\documentclass[a4paper,10pt]{article}
\usepackage{amssymb}
\usepackage{amsmath}
\usepackage{amsfonts}
\usepackage{graphicx}
\usepackage{color}
\usepackage{xspace}
\usepackage{color}
\usepackage{float}
\usepackage{hyperref}
\usepackage[numbers,sort&compress]{natbib}
\usepackage{makeidx}
\usepackage{braket}
\usepackage{dsfont}
\usepackage{tabularx}
\usepackage{authblk}
\usepackage[margin=1.5in]{geometry}
\makeindex

\usepackage{myoptions}

\graphicspath{{./Images/}}
\title{Exploring the Membrane Theory of Entanglement Dynamics}
\author{M\'ark Mezei}
\affil{
Simons Center for Geometry and Physics, SUNY, Stony Brook, NY 11794, USA}
\author{Julio Virrueta}
\affil{
C.N. Yang Institute for Theoretical Physics, SUNY, Stony Brook, NY 11794, USA}
\begin{document}

\maketitle
\begin{abstract}
\noindent Recently an effective membrane theory valid in a ``hydrodynamic limit'' was proposed to describe entanglement dynamics of chaotic systems based on results in random quantum circuits and holographic gauge theories. In this paper, we show that this theory is robust under a large set of generalizations. In generic quench protocols we find that the membrane couples geometrically to hydrodynamics, joining quenches are captured by branes in the effective theory, and the entanglement of time evolved local operators can be computed by probing a time fold geometry with the membrane. We also demonstrate that the structure of the effective theory does not change under finite coupling corrections holographically dual to higher derivative gravity and that subleading orders in the hydrodynamic expansion can be incorporated by including higher derivative terms in the effective theory.

\end{abstract}
\tableofcontents

\section{Introduction}\label{sec:Intro}

Entanglement dynamics in systems out of equilibrium is a rich phenomenon that has been studied in many areas of physics \cite{Kaufman794, Calabrese:2005in, Mezei:2016wfz, Mezei:2016zxg, Liu:2013iza, Liu:2013qca, Hartman:2013qma, Zhou:2018myl, Jonay:2018yei, Nahum:2018pcr}. The quasiparticle model \cite{Calabrese:2005in,Calabrese:2007rg,Casini:2015zua} provides an intuitive way to understand the general characteristics of entanglement dynamics, however, while it is an accurate description of integrable theories \cite{Calabrese:2016xau,Cotler:2016acd}, it does not capture all the properties present in a chaotic system \cite{Asplund:2015eha,Casini:2015zua,Liu:2013qca, Liu:2013iza}.

Recently, a new effective model has been proposed: the membrane theory. First discovered in the context of unitary random evolution in two dimensions \cite{Jonay:2018yei}, it has been generalized to arbitrary dimensions using the holographic correspondence \cite{Mezei:2018jco}. In this model, the problem of computing entanglement entropy is translated into the problem of computing the ``energy'' of a minimal timelike codimension-1 membrane of angle dependent tension $\mathcal{E}(v)$, which connects two faces of a slab of d-dimensional Minkowski spacetime of height T, where $v$ is a local velocity of the membrane,  the time component of the unit normal vector of the membrane. The entanglement entropy is then:
\begin{equation}
\label{membraneoriginal}
S = s_{\text{th}}\int d^{d-1}y\sqrt{-\ga}\, \frac{\mathcal{E}(v)}{\sqrt{1-v^2}},
\end{equation}
where $s_{\text{th}}$  is the thermal entropy density, $y^i$ are the coordinates  and $\ga_{ij}$ is  the induced metric on the membrane world volume.  

This theory is expected to accurately describe the dynamics of entanglement in the scaling regime: $R, T \gg \beta$, where $R$ is the characteristic size of the entangling region, $T$ is the time elapsed since the quench and $\beta$ is the inverse temperature of the system at equilibrium. Furthermore, it has been shown to successfully capture many of the important properties of the dynamics of operators in chaotic systems, specially the relation between entanglement dynamics and operator spreading \cite{Mezei:2016wfz}. In this paper we will show that the membrane theory holds for a broad range of generalizations. 

In \cite{Mezei:2018jco}, the initial excited state was produced by a global quench which preserves translational invariance. We will show that the membrane theory is valid for more general initial states, in particular those that do not exhibit translational invariance. At late enough time, once the system has reached local equilibrium, the coarse-grained dynamics can be described in terms of the diffusive transport of a few conserved quantities \cite{Lux_2014}. If the excitations of the system are characterized by sufficiently long wavelengths, we can study this diffusive transport in terms of a hydrodynamic expansion. 

Hydrodynamic systems possess a well-established holographic description \cite{Rangamani:2009xk, Hubeny:2010wp, Bhattacharyya:2008jc}, for which the long wavelength approximation is seen as a gradient expansion for an inhomogeneous black brane solutions. We use this fluid/gravity correspondence to compute the holographic entanglement entropy and show that in the scaling limit it obeys the membrane theory prescription. We also provide further evidence for the validity of the membrane theory by studying other inhomogeneous setups: the joining quench for semi-infinite systems separately in thermal equilibrium is described by the membrane theory with an added brane on which membranes can end, and the 
 the entanglement entropy of time evolved local operators is computed by a membrane living in a double cone geometry representing the footprint of the growing operator \cite{Roberts:2014isa}.

A key ingredient in the derivation of the membrane theory for higher dimensions is the holographic correspondence. Hence it is of interest to see if this effective theory holds under generalizations of the simplest holographic setup in terms of Einstein gravity. We do this by considering higher derivative gravity theories, the simplest example being Gauss-Bonnet gravity, and the most complicated explicitly analyzed is the general four derivative gravity. In addition to correcting to the geometry of the spacetime, these new terms also modify the holographic entanglement entropy functional \cite{Camps:2013zua,Dong:2013qoa}. Once again we show that the membrane theory is robust under the deformation of the bulk gravitational action,  and data read off from the membrane tension function $\mathcal{E}(v)$ reproduces previous results about entanglement growth in higher derivative gravity obtained in \cite{Mezei:2016wfz}. We also show how to incorporate subleading corrections in $\beta/R$ into the theory. These should be thought of as analogs of higher gradient terms in hydrodynamics.

We organize this paper as follows: in Sec.~\ref{sec:Review} we revisit the original derivation of the membrane theory for global quenches, setting the basic simplifications achieved by the scaling limit. In Sec.~\ref{sec:Hydro} we generalize this to consider systems without translational invariance, described in terms of the Fluid/Gravity correspondence, we show that the basic description in terms of a codimension-1 membrane still holds but we need to generalize it to include a coupling  the membrane to the fluid. We follow this by introducing a specific quench, the thermal joining quench and analyzing its entropy in the scaling limit. In Sec.~\ref{sec:OpGrowth} we consider one more case, corresponding to a state created by the action of a particular operator, which is dual to the shock-wave geometry. We then consider holographic entanglement entropy in theories with higher derivatives in Sec.~\ref{sec:HD}. We discuss next-to-leading order corrections in the scaling limit in Sec.~\ref{sec:NLO}. We end with a summary, discussion and open questions in Sec.~\ref{sec:Summary}.

\section{Membrane theory for global quenches\label{sec:Review}}

\begin{center}
\begin{figure}[H]
\centering
\includegraphics[scale=0.8]{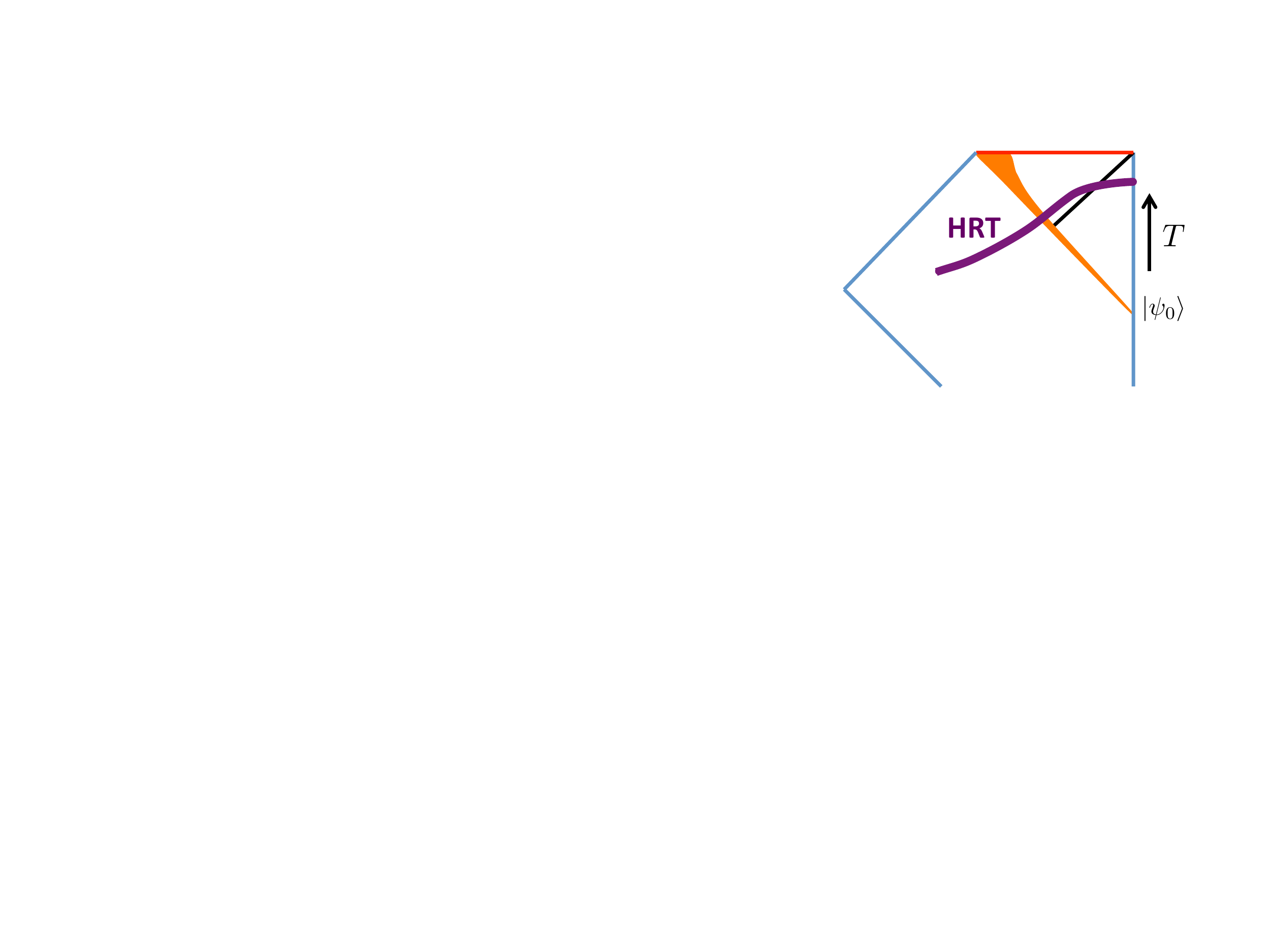}
\caption{Penrose diagram of a spacetime dual to a quench. Before the quench we have pure AdS, the infalling matter shell is colored orange, where the spacetime is strongly time dependent, and the spacetime subsequently settles to a static black brane.  The horizon is a diagonal black line, the singularity is a red line, the Poincare horizon and the AdS boundary are drawn by blue, while the HRT surface computing the entropy of half space is a purple curve.
\label{fig:Penrose}}
\end{figure}
\end{center}

Let us first review the original derivation of the membrane theory, as first presented in \cite{Mezei:2018jco}. As was noticed in the Introduction, the system is taken to be on a state $\ket{\psi} = e^{-iHT}\ket{\psi_0}$, where $\ket{\psi_0}$ is a highly excited short-range entangled state, prepared by a global quench. 
This configuration is dual to a dynamical spacetime modeling black brane formation from collapse, represented by the Penrose diagram Fig.~\ref{fig:Penrose}. We are interested on the entanglement entropy for a spatial subregion with characteristic size $R$, which can be computed holographically by the usual prescription in terms of extremal surfaces \cite{Hubeny:2007xt}.
 In order to implement the limit $R,T\gg \beta$, we follow \cite{Agon:2019qgh} and introduce a book keeping parameter $\Lambda\gg 1$, performing the transformation: $R,T \rightarrow \Lambda R, \Lambda T$. We are interested only on the leading $\Lambda^{d-1}$ contributions to the entanglement entropy. 

A key observation is that, in this approximation, only the part of the HRT surface that lies behind the horizon of the final black brane geometry contributes to the entropy\cite{Mezei:2016wfz, Mezei:2018jco}. The other pieces of the surface contribute only to order $\Lambda^{d-2}$, this includes in particular the usual area law contribution for the entanglement entropy of the ground state.

The most general static black brane geometry can be written in infalling coordinates as:
\begin{equation}
ds^2 = \frac{1}{z^2}\left(-a(z)dt^2 -\frac{2}{b(z)}dtdz+dr^2 + r^2 d\Omega_{d-2}^2\right),
\end{equation}
and a codimention-2 surface can be defined by the embedding $z(t,\Omega)$ and $r(t,\Omega)$, where $\Omega$ stands for the collective coordinates in $S^{d-2}$. The entanglement entropy is then computed by extremizing the area functional:\footnote{In the following expressions, all products involving the angular coordinates are taken using the metric on $S^{d-2}$ and $\dot{f} = \partial_t f$.}
\begin{equation}
\label{AreaFunct}
\begin{aligned}
S &= \frac{1}{4G_N}\int dtd\Omega \frac{r^{d-2}}{z^{d-1}}\sqrt{Q},\\
Q&= \left[\dot{r}^2 - \left(1+\frac{\left(\partial_{\Omega}r\right)^2}{r^2}a(z)\right)\right] + \frac{2}{b(z)}\left[\dot{r}\frac{\partial_{\Omega}r\cdot\partial_{\Omega}z }{r^2} - \left(1+\frac{\left(\partial_{\Omega}r\right)^2}{r^2}\dot{z}\right)\right] - \frac{\left(\partial_{\Omega}z\right)^2}{r^2 b(z)^2}.
\end{aligned}
\end{equation}

We implement the rescaling:
\begin{equation}
\label{CoordScaling}
t\rightarrow \Lambda t, \hspace{0.5cm} r \rightarrow \Lambda r, \hspace{0.5cm} \Omega\rightarrow\Omega, \hspace{0.5cm} z\rightarrow z.
\end{equation}

At leading order in $1/\Lambda$, the equation of motion for $z(t,\Omega)$ becomes algebraic and it is given by:
\begin{equation}
\label{projectionz}
v^2 \equiv \frac{\dot{r}^2}{1+\frac{(\partial_{\Omega}r)^2}{r^2}} = a(z) - \frac{z a'(z)}{2(d-1)} \equiv c(z),
\end{equation}
from where we can solve $z(t,\Omega) = c^{-1}\left(v^2(t,\Omega)\right)$ and rewrite the area functional as:
\begin{equation}
\begin{aligned}
S &= s_{th}\int dt d\Omega r^{d-2}\sqrt{1+\frac{(\partial_{\Omega}r)^2}{r^2}}\mathcal{E}(v),\\
\mathcal{E}(v) &=\left. \sqrt{\frac{-a'(z)}{2(d-1)z^{2d-3}}}\right|_{z=c^{-1}(v^2)},
\end{aligned}
\end{equation}
where we introduced $s_{th} = \frac{1}{4G_N}$ and set $\Lambda=1$.

\begin{center}
\begin{figure}[!h]
\centering
\includegraphics[scale=0.4]{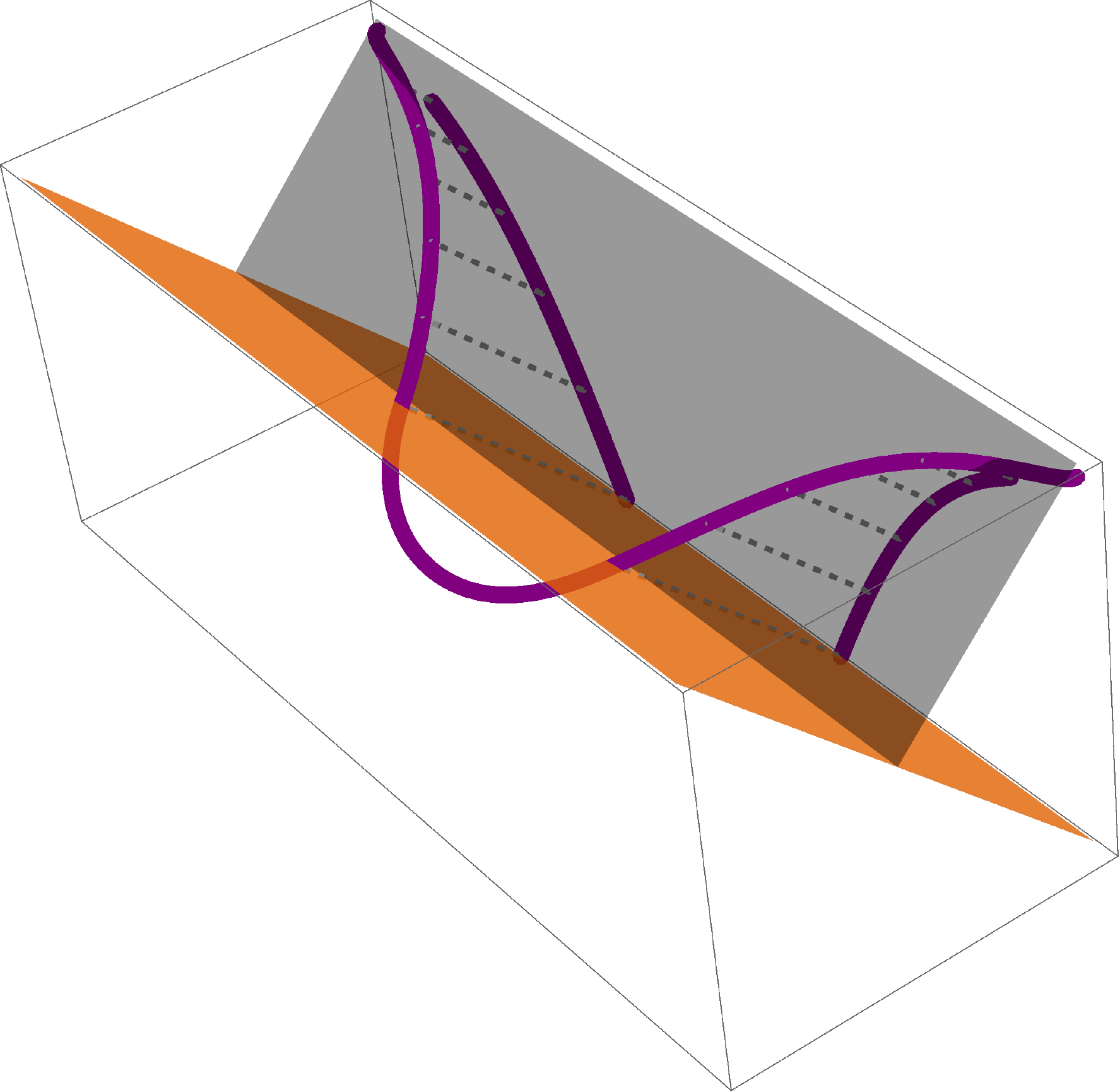}
\caption{Projection of the HRT surface into the membrane picture. Only the region between the infalling matter (Orange) and the horizon (Gray) contribute to leading order in the scaling limit.} 
\label{fig:Penrose2}
\end{figure}
\end{center}

We see that the problem of obtaining the HRT surfaces translates, in this limit, to the problem of minimizing a codimension-1 surface, or membrane, extending along the interval $[0,T]$ and with the boundary condition that, at $t=T$, it is equal to the codimension-2 entangling region,\footnote{At this boundary the relation between $v$ and $z$ \eqref{projectionz} breaks down. This is because the HRT surface stops obeying the scaling Ansatz exactly at this point. } while on the $t=0$  surface representing the short range entangled initial state  it ends perpendicularly. The information about the original holographic set up is encoded on the \textit{membrane tension} $\mathcal{E}(v)$ and by solving the membrane problem one can reconstruct the full HRT surface, to leading order, using the map \eqref{projectionz}. The projection of the HRT surface into the membrane can be seen in Fig.~\ref{fig:Penrose2}.

There is one more ingredient that is required to make the membrane theory match the HRT prescription. We also have to allow for horizontal membranes (that formally give $v=\infty$) to capture the saturation of entropy  \cite{Mezei:2018jco}. These do not have to extend down to the $t=0$ boundary, and give $S=s_\text{th} \vol(A)$ when they are the minimal action membrane. 

The membrane tension $\mathcal{E}(v)$ obeys a series of consistency conditions that encode information about operator spreading \cite{Jonay:2018yei}, namely, it is a positive even function, monotonically increasing and convex for $0\leq v <1$. It diverges as $v\rightarrow 1$ and certain important values are \cite{Mezei:2018jco}:
\begin{equation}
\label{velocities}
\mathcal{E}(0) = v_E, \hspace{0.5cm} \mathcal{E}(v_B) = v_B, \hspace{0.5cm} \mathcal{E}'(v_B) = 1.
\end{equation}

In the following sections we will use these relations to compute corrections to the butterfly and entanglement velocities as well as consistency checks for the membrane theory.

\section{Entanglement dynamics in the hydrodynamic limit}\label{sec:Hydro}

A generalization, with respect to the previous picture, is to take more general initial states. In the scaling limit we are interested in studying systems out of equilibrium at times $T\gg \beta$. In this regime, generic states without translational invariance can be described in terms of a hydrodynamic expansion in the low energy/long wavelength limit. Let us then begin this section by reviewing the fundamental ideas of the fluid/gravity correspondence 
(see \cite{Bhattacharyya:2008jc, Hubeny:2010wp, Rangamani:2009xk} for details). 

The degrees of freedom of the effective hydrodynamic description are encoded in the stress tensor $T_{ab}$ and conserved current $J_{a}$, which in turn can be expressed in terms of a velocity field $u_{a}$ and certain scalar functions, such as the pressure, temperature, energy density, and conserved charges. The dynamics of these quantities is determined by the conservation equations and the equation of state of the medium.

We will consider a theory in $d-$dimensional Minkowski spacetime, for simplicity we assume the theories has no global symmetries and we focus only in the dynamics of the stress tensor. The hydrodynamic regime is organized in terms of a gradient expansion, at leading order we have an ideal fluid, for which:
\begin{equation}
T_{ab}^{(0)} = \rho u_{a}u_{b}+ p P_{ab},
\end{equation}
where $\rho$ is the energy density, $p$ is the pressure, and $P_{ab} =\eta_{ab}+u_a u_{b}$ is a projection operator into the plane orthogonal to the velocity field, we normalized the later as $u_a u^a =-1$.

To next order in the gradient expansion, the stress tensor receives contributions from derivatives of the velocity field, these can be organized by their transformation under the Lorentz group as:
\begin{equation}
\label{firstorder}
\begin{aligned}
\theta &= \partial_a u^a, \\
a_a &= u^b \partial_b u_a,\\
\sigma^{ab} &= \partial^{(a}u^{b)} + u^{(a}a^{b)} - \frac{1}{d-1}\theta P^{ab},\\
w^{ab} &= \partial^{[a}u^{b]} + u^{[a}a^{b]}.
\end{aligned}
\end{equation}

By symmetry considerations these quantities contribute to the stress tensor as:
\begin{equation}
T_{ab}^{(1)} = -2 \eta \sigma_{ab}- \zeta \theta P_{ab},
\end{equation}
where $\eta$ and $\zeta$ are transport coefficients, known as the shear and bulk viscosities, respectively. 

To all orders in the hydrodynamics expansion, the stress tensor possesses the same structure: we will have a series of tensors, formed out of the velocity field and its derivatives, whose contribution to the stress tensor is characterized by a series of transport coefficients, which encode the specific details of the underlying theory. We will see that the membrane theory follows a similar structure. Just as these transport coefficients depend on the temperature, the angle dependent membrane tension ${\cal E}(v)$ can be regarded as a generalized transport coefficient of the membrane effective theory: the form of the theory is universal, but ${\cal E}(v)$ is specific to a theory.  In this section we will introduce new membrane theory transport coefficients that determine the entropy dynamics in inhomogeneous states.

In order to construct the holographic dual of this hydrodynamic theory, one considers first a boosted AdS-Schwartzschild black brane, characterized by a constant timelike vector $u_a$:
\begin{equation}
ds^2 = \frac{1}{z^2}\left(2u_a dx^a dz + \left(\eta_{ab} + \left(1-a\left(d z/4\pi T\right)\right)u_a u_b\right)dx^a dx^b\right),
\end{equation}
which is a vacuum solution of the Einstein equations with negative cosmological constant and $a(\zeta)=1-\zeta^d$.\footnote{Note that in Sec.~\ref{sec:Review} we  set $T=d/(4\pi)$, hence had $\zeta=z$.} We then promote the vector $u_a$ and the temperature $T$ to be functions of $x=(t,\vec{x})$. The result is no longer a vacuum solution, but one can systematically correct the metric in the form $g_{\mu\nu} = g_{\mu\nu}^{(0)} + g_{\mu\nu}^{(1)}+...$, organized as a gradient expansion in terms of derivatives of $u_a(x)$ and $T(x)$, to obtain a solution. As showed first in \cite{Bhattacharyya:2008jc}, this perturbative expansion of the Einstein equations reproduces order by order the hydrodynamic expansion of the dual field theory.

The most general solution to the perturbative equations takes the form\cite{Rangamani:2009xk}:
\begin{equation}
\label{hydrometric}
ds^2 = \frac{1}{z^2}\left\{2u_a(x) dx^a dz + \left[G_{ab}(z,x)-2u_a(x) B_b(z,x)\right]dx^adx^b \right\},
\end{equation}
where the functions $G_{ab}(z,x)$ and $B_b(z,x)$ are determined order by order in the gradient expansion. To leading non-trivial order:
\begin{equation}
\begin{aligned}
B_{a}(z,x) &=  \frac{1}{2}a\left(d z/4\pi T\right)u_a + z \mathcal{A}_a,\\
G_{ab}(z,x) &= P_{ab} +\frac{8\pi T}{d} g_1\left(d z/4\pi T\right) \sigma_{ab},
\end{aligned}
\end{equation}
where $\mathcal{A}_a = a_a-\frac{1}{d-1}\theta u_a$. The function  $g_1(\zeta)$ depends only in the dimension of spacetime and is known in integral form:
\begin{equation}
g_1(\zeta) = \int_\zeta^{\infty} dy \frac{y^{d-1}-1}{y(y^d-1)}.
\end{equation}

The local entropy density of the fluid is also coordinate dependent and it is given, up to second order, by:
\begin{equation}
\label{entropy}
s(x) = \frac{1}{4 G_N}\left(\frac{4\pi T(x)}{d}\right)^{d-1}.
\end{equation}

\subsection{Entanglement entropy and the scaling limit}
We want to consider the scaling limit by performing the rescaling \eqref{CoordScaling}. In doing this, one must also be mindful of the way the metric rescales, and hence must specify the particular scaling for each of the quantities in the gradient expansion. For the velocity field $u_a(x)$ we have:
\begin{equation}
u_r \rightarrow u_r, \hspace{1cm} u_t\rightarrow u_t, \hspace{1cm} u_{\Omega} \rightarrow \Lambda u_{\Omega},
\end{equation}
which preserves the normalization condition $u_a u^a=-1$ in the new, rescaled, coordinates. We observe that due to the explicit $r^2$ factor in the sphere metric, the angular components are rescaled, while other components are not scaled. This pattern holds true for the different tensors constructed out of the velocity field and its derivatives. To first order in the gradient expansion the quantities that require rescaling are:
\begin{equation}
\begin{aligned}
a_{\Omega} &\rightarrow \Lambda a_{\Omega},\\
\sigma^{r\Omega}, \sigma^{t\Omega} &\rightarrow \frac{1}{\Lambda}(\sigma^{r\Omega}, \sigma^{t\Omega}),\\
\sigma^{\Omega\Omega} &\rightarrow \frac{1}{\Lambda^2}\sigma^{\Omega\Omega},
\end{aligned}
\end{equation}
and the antisymmetric tensor $\omega^{ab}$ follows the same rescaling as $\sigma^{ab}$.

We then compute the holographic entanglement entropy for the same set up as in Sec.~\ref{sec:Review}. The area functional is given, to zeroth order in the gradient expansion, by:
\begin{equation}
\begin{aligned}
S &= \frac{\Lambda^{d-1}}{4G_N}\int dt d\Omega \frac{r^{d-2}}{z^{d-1}}\sqrt{Q},\\
Q &= \dot{r}^2- a\left(d z/4\pi T\right)\left(1+\frac{(\partial_{\Omega}r)^2}{r^2}\right) +\left(1-a\left(d z/4\pi T\right)\right) \left(1+\frac{(\partial_{\Omega}r)^2}{r^2}\right)w^2,\\
w^2 &= \frac{(u_{r}^2+\frac{u_{\Omega}^2}{r^2})\dot{r}^2+(u_{r}^2+\frac{u_{\Omega}}{r^2}\frac{(\partial_{\Omega}r)^2}{r^2})-2\frac{u_{\Omega}\cdot\partial_{\Omega}r}{r^2}(u_{t}\dot{r}+u_{r})+2u_{t}u_{r}\dot{r}}{1+\frac{(\partial_{\Omega}r)^2}{r^2}},
\end{aligned}
\end{equation}
where all the information about the fluid is contained in the function $w$. Note that in the homogenous setting $w=0$, and we recover the area functional \eqref{AreaFunct} in the scaling limit.
This expression does not look very enlighting, however it posses a key property, the functional does not contain derivatives of $z(t,\Omega)$, so the corresponding equation of motion is algebraic.

We can further simplify the previous expression by writing it in terms of contractions of tensors characterizing the fluid and the membrane. To see this we introduce the vector, in $d-$dimensional Minkowski spacetime, normal to the entangling region:
\begin{equation}
n_{a} = \frac{1}{\sqrt{1+\frac{(\partial_{\Omega}r)^2}{r^2}-\dot{r}^2}}(-\dot{r},1,-\partial_{\Omega}r),
\end{equation}
where we use the $(t,r,\Omega)$ ordering of coordinates. We can then rewrite the function Q as:\footnote{In the case where $(u_{t},u_{r},u_{\Omega})=(-1,0,0)$, we have $v^2 \rightarrow \frac{\dot{r}^2}{1+\frac{(\partial_{\Omega}r)^2}{r^2}}$, recovering the translational invariant case. }:
\begin{equation}
\label{velocityfluid}
\begin{aligned}
Q &=\frac{ \left(1+\frac{(\partial_{\Omega}r)^2}{r^2}-\dot{r}^2\right)}{(1-v^2)}
\left(v^2-a\left(d z/4\pi T\right)\right),\\
v^2 &= \frac{(n\cdot u)^2}{1+(n\cdot u)^2},
\end{aligned}
\end{equation}
where we notice only the second factor has a $z$ dependence. 

Since the temperature $T(x)$ is a non-trivial function of the boundary coordinates, it is convenient to introduce a new variable $\zeta = \frac{d z}{4\pi T(x)}$. The algebraic equation of motion for the embedding function $\zeta(t,\Omega)$ is then
\begin{equation}
\label{zEoM}
v^2 = a(\zeta) - \frac{\zeta a'(\zeta)}{2(d-1)}\equiv c(\zeta),
\end{equation}
which can be solved to obtain $\zeta$ as a function of $v^2$. Using this equation we can rewrite the area functional as
\begin{equation}
\begin{aligned}
S &=\Lambda^{d-1} \int d^{d-1}y\sqrt{-\ga}\, \frac{s(x)\mathcal{E}(v)}{\sqrt{1-v^2}},\\
\mathcal{E}(v) &= \left. \sqrt{\frac{-a'(\zeta)}{2(d-1)\zeta^{2d-3}}}\right|_{\zeta=c^{-1}(v^2)},
\end{aligned}
\end{equation}
where in spherical coordinates $d^{d-1}y\sqrt{-\ga} =r^{d-2} \sqrt{1+\frac{(\partial_{\Omega}r)^2}{r^2}-\dot{r}^2}dt d\Omega$ is the area element for a codimension one surface in Minkowski spacetime, characterized by the embedding $r=r(t,\Omega)$. We used \eqref{entropy} to rewrite all $T(x)$ dependence in terms of the entropy density.

We see that this expression is precisely the membrane theory prescription \eqref{membraneoriginal}, with the two generalization being that now the entropy density is a function of the coordinates and the velocity $v$ is measured with respect to $u$ instead of $\hat{t}$. This is a beautiful minimal coupling of the membrane to the fluid. Next we consider the leading corrections to this action in the fluid gradient expansion, which induces nonminimal couplings between the fluid and the membrane.

As before, we compute the area functional using the metric \eqref{hydrometric} and the rescaled embedding \eqref{CoordScaling}, and we express the result in terms of invariant products of the form $(n\cdot u)$ and products of the normal vector with the quantities defined in \eqref{firstorder}, we have that:
\begin{equation}
\begin{aligned}
Q &=\frac{ \left(1+\frac{(\partial_{\Omega}r)^2}{r^2}-\dot{r}^2\right)}{(1-v^2)}
\left(v^2-a(\zeta) - \zeta \frac{8\pi}{d}T(x)\left(\frac{\left(\mathcal{A}\cdot n\right)(n\cdot u)-\mathcal{A}\cdot u}{1+(n\cdot u)^2}\right)\right.\\
&+\left.  \frac{8\pi}{d}T(x) g_1(\zeta)a(\zeta)\frac{\sigma_{ab}n^a n^b}{1+(n\cdot u)^2}\right).
\end{aligned}
\end{equation}
We notice that there is no contribution from the antisymmetric tensor $\omega_{ab}$, this is due to symmetry: we cannot form an invariant product involving only that tensor and $n_a$. The particular form of the first correction term is due to conformal symmetry, instead of the quantities $a_a = u^b\partial_b u_a$ and $\theta = \partial^a u_a$ appearing separately, they arrange themselves into the Weyl connection $\mathcal{A}_a = a_a - \frac{\theta}{d-1}u_a$. 

It is convenient to define new $\zeta$ independent variables, which encode the higher order corrections:
\begin{equation}
\begin{aligned}
\mathcal{Q}_1 &=  \frac{8\pi}{d}\left(\frac{\left(\mathcal{A}\cdot n\right)(n\cdot u)-\mathcal{A}\cdot u}{1+(n\cdot u)^2}\right)T(x),\\
\mathcal{Q}_2 &=  \frac{8\pi}{d} \frac{\sigma_{ab}n^a n^b}{1+(n\cdot u)^2}T(x).
\end{aligned}
\end{equation}
The algebraic equation of motion for $\zeta(x)$ is then
\begin{equation}
\label{aom2}
v^2 = a(\zeta) - \frac{\zeta a'(\zeta)}{2(d-1)} +  \frac{2d-3}{2(d-1)} \zeta \mathcal{Q}_1 +\left(\frac{\zeta g_1(\zeta)a'(\zeta)}{2(d-1)}-a(\zeta)g_1(\zeta) + \frac{\zeta a(\zeta)g_1'(\zeta)}{2(d-1)}\right) \mathcal{Q}_2.
\end{equation}

Unlike the zeroth order case, we cannot write this simply in the form $v^2 = c(\zeta)$, instead we must solve this equation order by order in the gradient expansion:
\begin{equation}
\label{sysEoM}
\begin{aligned}
c(\zeta_{(0)}) = a(\zeta_{(0)}) - \frac{\xi_{(0)} a'(\zeta_{(0)})}{2(d-1)} &= v^2,  \\
\zeta_{(1)} &= F_{(1)}(v,\partial v),\\
\zeta_{(2)} &= F_{(2)}(v,\partial v,\partial^2 v),\\
&\vdots
\end{aligned}
\end{equation}
where $\partial^n v$ denotes the nth-order corrections to the fluid/gravity metric, contracted with the normal vector $n_a$. For instance:
\begin{equation}
 F_{(1)}=\frac{(2d-3)\mathcal{Q}_1+ \left(g_1(\zeta_{(0)})a'(\zeta_{(0)})+a(\zeta_{(0)})\left(g_1'(\zeta_{(0)})-2(d-1)\frac{g_1(\zeta_{(0)})}{\zeta_{(0)}}\right)\right)\mathcal{Q}_{2}}{a''(\zeta_{(0)})-(2d-3)\frac{a'(\zeta_{(0)})}{\zeta_{(0)}}},
\end{equation}
where $\zeta_{(0)} = c^{-1}(v^2)$.

The equation \eqref{aom2} can then be use to write the area functional as
\begin{equation}
\label{hydroResult}
\begin{aligned}
S &= \Lambda^{d-1}\int d^{d-1}y\sqrt{-\ga}\, \frac{s(x)\mathcal{E}(v)}{\sqrt{1-v^2}},\\
\mathcal{E}(v) &=\mathcal{E}^{(0)}(v)\left(1+q_1(v)\mathcal{Q}_1+q_2(v) \mathcal{Q}_{2}\right),\\
&=\mathcal{E}^{(0)}\left(1 + \frac{8\pi T(x)}{d(1+(n\cdot u)^2)}\left(q_1 \left( \left(\mathcal{A}\cdot n\right)(n\cdot u)-\mathcal{A}\cdot u\right)+q_2\sigma_{ab}n^a n^b\right)\right),
\end{aligned}
\end{equation}
where
\begin{equation}
\begin{aligned}
\mathcal{E}^{(0)}(v) &= \left. \sqrt{\frac{-a'(\zeta)}{2(d-1)\zeta^{2d-3}}}\right|_{\zeta_{(0)}=c^{-1}(v^2)},\\
q_1(v) &= \left. \frac{d-1}{a'(\zeta_{(0)})}\right|_{\zeta_{(0)}=c^{-1}(v^2)},\\
q_2(v) &= - \left.\frac{d-1}{\zeta_{(0)}a'(\zeta_{(0)})}a(\zeta_{(0)})g_1(\zeta_{(0)}) \right|_{\zeta_{(0)}=c^{-1}(v^2)}.
\end{aligned}
\end{equation}
We see that the dissipative corrections modify the tension function $\mathcal{E}(v)$ but they still can be taken into account within the framework of the membrane theory. Furthermore the corrections appear as we expected, this is, as tensor structures made out of  invariant products of the vector $n_a$ characterizing the membrane and the different tensors describing the fluid dynamics. 

Even though \eqref{hydroResult} was derived using holography, one may expect that the membrane theory holds beyond that framework; one just considers a membrane in Minkowski spacetime, non-minimally coupled to a fluid with coefficients dependent in the particular details of the theory. Similarly, although \eqref{hydroResult} was derived only to leading order in the dissipative corrections, it is easy to see that it holds for higher corrections, one just need to incorporate further tensor structures in \eqref{aom2} and solve the equation order by order in the gradient expansion, obtaining a result of the form:
\begin{equation}
\mathcal{E}(v) = \mathcal{E}^{(0)}(v)\left(1 + \sum_{i=1}^n\sum_I q_I^{(i)}\mathcal{Q}_I^{(i)}\right),
\end{equation}
where the first sum is over the gradient expansion, up to order $n$; while the second sum is over the different tensor structures, at a given order, allowed by symmetry.

We have determined the bulk Lagrangian of the membrane theory in an inhomogeneous quench. To make the problem well-defined, we have to specify the boundary conditions on the membrane. At $t=T$ the membrane is anchored on the subregion $A(T)$. The other boundary condition is specified by the quench protocol that created the state. One straightforward protocol would be to create a short range entangled state with a prescribed density of conserved charges at $t=0$. Strongly coupled chaotic systems are expected to loose the memory of the details of the initial state in a time of order $\beta$, which is dual to black brane quasinormal modes decaying in times of order $\beta$ (except for the hydrodynamic ones). Hence for $1\leq z\leq z_*$ the spacetime settles to a fluid/gravity metric in a short time of order $\beta$. The membrane theory  is insensitive to such short time details, hence the whole quench protocol is represented as a brane at $t=0$ on which the membrane can end perpendicularly. We can consider some variation on this setup: the quench could be implemented at different times at different spatial locations, resulting in a wavy membrane $t_\text{quench}(x)$. The initial state could also contain significant amount of entanglement, in which case the membrane action has to be supplemented by a boundary term \cite{Jonay:2018yei}, and as a result will 
obey different boundary conditions at $t=0$.

\section{Joining quench}\label{sec: Joining}

In the previous section we showed that, for a generic quench without translational invariance, the dynamics of entanglement entropy is characterized by the membrane theory coupled to a fluid. We did this by considering an effective hydrodynamic description of the quench after local equilibration. In this section we will consider a particular quench without translational invariance, study its time dependence in general, and show that it is accurately captured by the membrane theory in the scaling limit.

We consider two decoupled theories on half spaces that are separately thermalized. We can quench this system by coupling the two theories along their boundaries, as was discussed in the random circuit context in \cite{Jonay:2018yei}, by which we are heavily influenced. Joining quenches were studied in field theory in the vacuum state in many interesting papers \cite{Calabrese:2016xau,Ugajin:2013xxa, Shimaji:2018czt}.

In this section, we first construct a simple bottom up holographic model of this joining process. A simplifying feature of this model, is that it postulates the existence of an end of the world brane in the bulk theory that can be treated as a probe. Its presence results in a joining quench that involves no transport of energy, hence trivial hydrodynamics, but nontrivial entanglement dynamics. We find a simple membrane theory description of the entanglement dynamics. The results are in complete agreement with those obtained in the  random circuit context in \cite{Jonay:2018yei}, and complement it by providing a membrane theory spacetime picture. 

Next, we study conformal boundaries in CFT$_2$. The joining quench can be solved for using CFT techniques, and a bulk picture is obtained using mappings similar to those introduced in \cite{Roberts:2012aq}. The distinguishing feature of this quench model is that the joining is accompanied by an energy shock, i.e. the energy and entropy dynamics are coupled. A similar situation has been recently studied in the context of JT gravity \cite{Almheiri:2019psf, Almheiri:2019hni}.

While the entanglement entropy in the scaling regime agree with those obtained in the simplified model, the membrane theory description of the process seems considerably more complicated, and we only make initial steps towards deriving the membrane theory description of this particular joining quench. It would be very interesting to complete the derivation.

\subsection{A simple holographic model}

A holographic BCFT (on a half space) in the vacuum state has the dual gravitational description of a patch of AdS space ending on an end of the world brane \cite{Takayanagi:2011zk, Fujita:2011fp}. The brane satisfies boundary conditions:
\begin{equation}
K_{ab} = \left(K-T_\text{brane}\right)h_{ab},
\end{equation}
where $K_{ab}$ is the extrinsic curvature and $T_\text{brane}$ the tension. By demanding conformal boundary conditions on the boundary theory, the brane behaves as a $AdS_d$ foliation of $AdS_{d+1}$ and can be seen as an end of the world brane extending from the boundary of the half-plane into the bulk. The angle of the brane with the boundary is determined by the boundary conditions as:
\begin{equation}
T _\text{brane}= \frac{d-1}{L}\tanh\theta.
\end{equation}

In terms of field theory data, the angle is determined by the boundary entropy as $S_\text{bdy} = \frac{\theta}{4G_N}$. In the following, we choose $T_\text{brane}=\theta=0$ for simplicity. We expect that the results obtained in this special case should carry over to the $T_\text{brane}\neq 0$, where we would need to treat back reaction. The spectrum of brane tensions is given by the bulk string theory, and it would be interesting to find examples, where the end of the world brane can be approximately tensionless. 

Intuitively, a joining quench corresponds to taking two BCFTs and gluing the branes together so that they become a folded brane and letting the folded brane freely fall into the bulk \cite{Astaneh:2014fga, Ficnar:2013wba, Chesler:2008uy}.\footnote{This model for joining quench was first proposed by \cite{Astaneh:2014fga}, here we follow the same model but not their calculation of the entanglement entropy.} While this setup certainly requires UV regularization, a folded brane whose tip makes its closest approach to the AdS boundary at $t=0$ should be a good model of a joining quench. The trajectory of the tip of the folded brane in AdS follows a null geodesic \cite{Ficnar:2013wba} which we take to be given by $z=z_0+\abs{t}$, where $z_0\propto {T _\text{brane}/( G_N E _\text{brane})}$ \cite{Astaneh:2014fga, Ficnar:2013wba}, that we take to be finite in the tensionless limit.

\begin{center}
\begin{figure}[!h]
\centering
\includegraphics[scale=0.65]{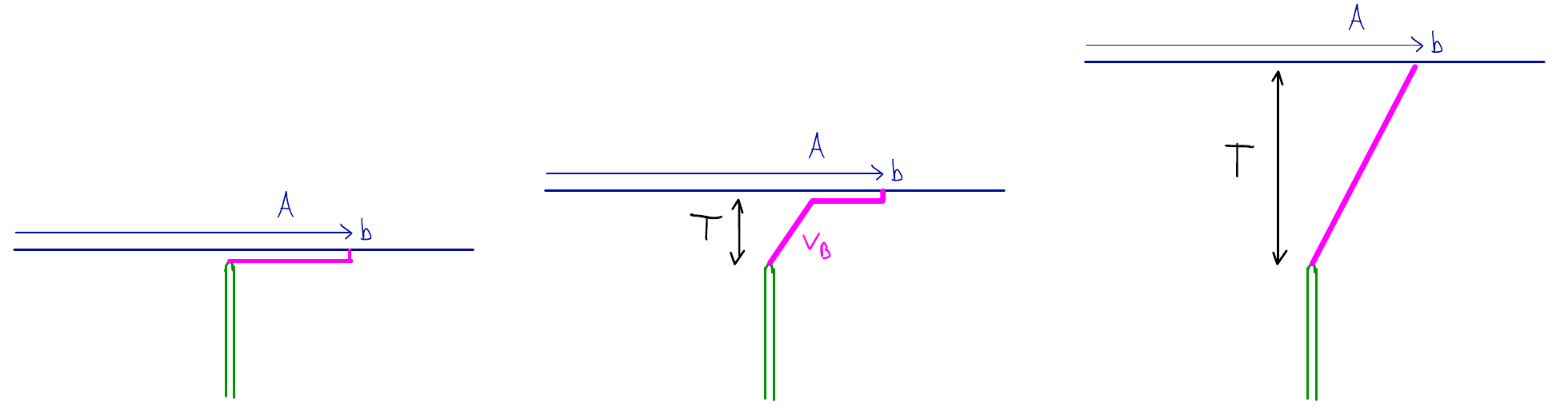}
\caption{The membrane theory description of the joining quench. The end of the world brane is the double half line from $t=0$ to $t=-\infty$, and membranes can end on it anywhere. We included the membranes for the half space $A(T)$ for $T=0$, a time $T<b/v_B$ and $T>b/v_B$. On the left figure the membrane is horizontal, and $S(T=0)= s_\text{th}A_\Sig b$. On the middle figure the membrane is composed of a horizontal piece and a ``light sheet'' of slope $v_B$. On the right figure the minimal membrane is a sheet of slope $b/T$. 
\label{fig:eow}}
\end{figure}
\end{center}

In the case, when the BCFTs are initially in thermal equilibrium, we have two black branes cut in half by end of the world branes, and we model the joining quench by a folded tensionless brane ``bouncing off'' the AdS boundary. Since there is no back reaction, the spacetime is that of a black brane for all times. In particular, there is no transport, and hydrodynamics is trivial. We just have a folded brane in the bulk, on which HRT surfaces can end, and this gives rise to the time dependence of entanglement entropy. In the hydrodynamic limit, all we have to do is include this end of the world brane in the membrane theory. Since the tip of the brane is on the $t=0$ infalling time plane, we get a codimension one end of the world brane also in membrane theory, extending from $t=0$ to $t=-\infty$, see Fig.~\ref{fig:eow}. The membrane can end anywhere on this brane.

\begin{center}
\begin{figure}[!h]
\centering
\includegraphics[scale=0.45]{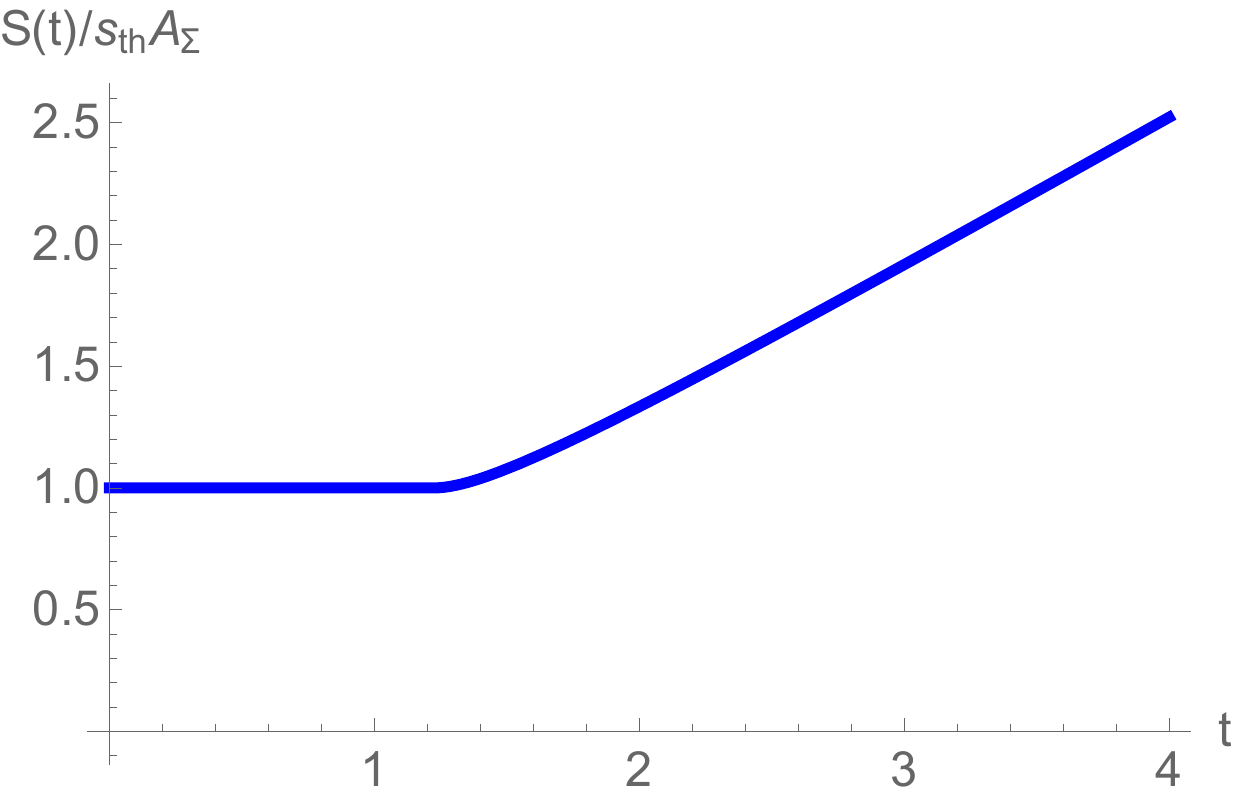}\hspace{0.7cm}
\includegraphics[scale=0.45]{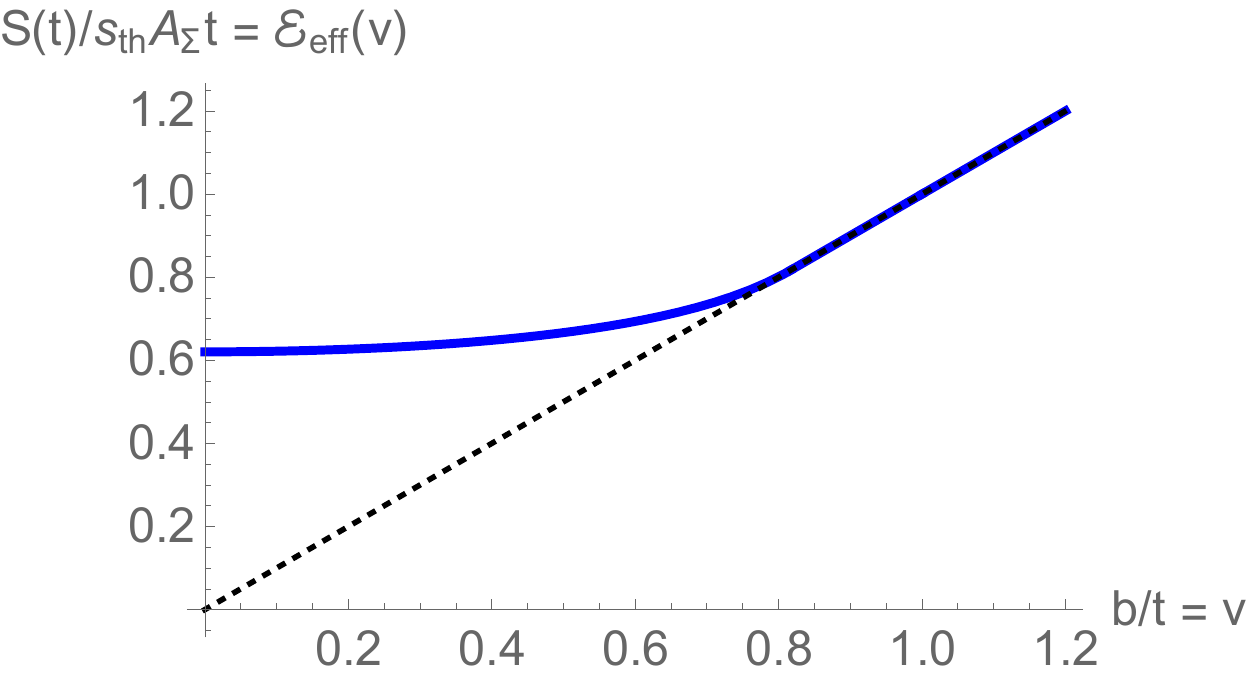}
\caption{{\bf Left:} $S(t)$ from \eqref{EntTime} plotted for $b=1$. The graph asymptotes to $S(T)\sim s_\text{th}A_\Sig\, v_E T$. {\bf Right:} If instead we fix time, and plot the entropy as a function of $v\equiv b/t$, we obtain directly the effective membrane tension function ${\cal E}_\text{eff}(v)$ \cite{Jonay:2018yei,Agon:2019qgh}, which agrees with ${\cal E}(v)$ for $v\leq v_B$, and is equal to $v$ for $v>v_B$. A black dotted  $45^\circ$ line is added to guide the eye.
\label{fig:halfspace}}
\end{figure}
\end{center}

We work out the example of the half space entangling surface defined by $x\in(-\infty,b)$. The membrane ends at the tip of the brane, and from what is explained on Fig~\ref{fig:eow} we get:
\es{EntTime}{
S(T)=s_\text{th}A_\Sig \begin{cases}
b\qquad &(v_B T<b)\,,\\
T\,{\cal E}\le(b\ov T\ri)&(v_B T\geq b)\,.
\end{cases}
}
Note that the function is continuous, because ${\cal E}\le(v_B\ri)=v_B$ \eqref{velocities}, see Fig.~\ref{fig:halfspace} for the graph of this function. Remarkably, from this graph we can read off the membrane tension function straightforwardly. Perhaps an even better visualization method comes from fixing $t$ and changing $b$, which directly maps out the membrane tension function  ${\cal E}\le(v\ri)$, see Fig.~\ref{fig:halfspace}. It would be interesting to work out the time evolution for other shapes. 

The same result \eqref{EntTime} has been obtained for random circuit models in \cite{Jonay:2018yei}. Here we provided a holographic derivation in a simple setup and gave a spacetime picture for the process in Fig.~\ref{fig:eow}. Below we analyze a CFT$_2$ joining quench, over which we have complete field theory (and holographic) control. A complication arises: the joining creates a shock of energy as in \cite{Almheiri:2019psf, Almheiri:2019hni}, which propagates through the system ballistically. While the time evolution of the entropy is identical to that in \eqref{EntTime}, the membrane theory description seems to be a lot more complicated in this case, as we explain in detail.

\subsection{An exactly solvable joining quench in CFT$_2$}

\subsubsection{Field theory computations}

For the case of a global quench, the initial state of the system can be approximated as a conformal boundary state $e^{-\frac{\beta}{4} H}\ket{B}$ \cite{Calabrese:2005in,Hartman:2013qma}. Such state is prepared by a path integral in Euclidean time, over a strip of width $\beta/2$. For theories with conformal invariance, the strip can be mapped into the half-plane. The time-dependence of correlation functions and entanglement entropy is then determined by general properties of conformal field theories with boundaries (BCFTs) \cite{Calabrese:2005in, Calabrese:2016xau}. 

In the absence of translational invariance, the initial state cannot be represented as a conformal boundary state. However,  it can still be prepared as a path integral over a Riemann surface, which can be conformally mapped into the half-plane \cite{Calabrese:2007mtj, Ugajin:2013xxa, Shimaji:2018czt}. For two-dimensional theories, it is convenient to introduce complex coordinates $w,\bar w$, with $w=x+i\tau$; the conformal map is then a biholomorphic transformation $w\rightarrow f(w)$. We can always choose the transformation such that the half-plane is given by $\operatorname{Re} f(w)\geq 0$.

As noticed earlier, a conformal field theory in the half-plane has a well-known holographic dual in terms of a section of $AdS$ spacetime, divided by a brane homologous to the boundary of the half-plane \cite{Takayanagi:2011zk, Fujita:2011fp}. Entanglement entropy can then be computed using the standard holographic prescription \cite{Hubeny:2007xt}. Due to the presence of a boundary, there are in general two possible extremal surfaces, a connected and a disconnected one, the holographic prescription then instructs us to take the surface with smaller area. 

For two-dimensional theories and a finite entangling regions $x\in[a,b]$, the two possible values of the holographic entanglement entropy are
\begin{equation}
\label{EE2d}
\begin{aligned}
S_{\text{con}} &= \frac{c}{6}\log\left(\frac{\left| f(w_1)-f(w_2)\right|^2}{\delta^2\left| f'(w_1)\right| \left| f'(w_2)\right|}\right),\\
S_{\text{disc}} &= \frac{c}{6}\log\left(\frac{4\operatorname{Re}f(w_1) \operatorname{Re} f(w_2)}{\delta^2\left| f'(w_1)\right| \left| f'(w_2)\right|}\right),
\end{aligned}
\end{equation}
where $\delta$ is the UV cutoff and we perform an analytic continuation $\tau=iT$, then
\begin{equation}
\begin{aligned}
w_1 &= a-T, \hspace{1cm} w_2 = b-T,\\
\bar{w}_1 &= a+T, \hspace{1cm} \bar{w}_2 = b+T.
\end{aligned}
\end{equation}

A specific example of a quench without translational invariance is the joining quench, where two semi-infinite systems are prepared on their respective ground states and joined together at time $T=0$, producing an excited state of the full Hamiltonian \cite{Ugajin:2013xxa, Shimaji:2018czt}, which however quickly settles back to the vacuum, and any subregion has subextensive entropy. In order to make contact with the regime of applicability of the membrane theory, we will modify this quench protocol: instead of the ground state, the semi-infinite systems will be prepared in thermal equilibrium. We refer to this as the thermal joining quench.

In the original joining quench, the initial state is prepared as a path integral in Euclidean time over the whole plane, except for branch cuts along $\tau\in(-\infty,-\epsilon]\cup[\epsilon,\infty)$ at $x=0$. In order to take the two systems to be in thermal equilibrium we compactify the Euclidean time direction on a circle of length $\beta$, the path integral is then over the thermal cylinder, except for a cut at $x=0$, along $\tau\in[-\beta/2,-\epsilon]\cup[\epsilon,\beta/2]$. 

Considering the transformations:
\es{ConfMap}{
G(w) &= -\left(\left(e^{\frac{2\pi}{\beta}w}+1\right)^{-1}-\frac{1}{2}\right),\\
F(w) &= w+\sqrt{w^2+\left(\frac{\pi \epsilon}{2\beta}\right)},
}
where the transformation $G(w)$ maps the cylinder to the plane by the usual exponential map and then translates the cut so that it coincides with the cut in the vacuum case, the second transformation $F(w)$ then maps this to the half-plane. We can then map the cut cylinder to the half-plane by the composite transformation $w\rightarrow f(w) = F(G(w))$. 

Using this conformal map in the general formula \eqref{EE2d} we obtain the entanglement entropy for the joining thermal quench. The result is dependent in the position of the entangling region with respect to the joining point at $x=0$ and the value of $T$. Without lost of generality we assume that $b>0$ and $|a|<b$, then we have two cases, depending on the sign of $a$. 

For the case $0<a<b$ we have:
\begin{equation}
S_{\text{con}} = 
\begin{cases}
\frac{c}{3}\log\left(\frac{\beta}{\pi \delta}\sinh\left(\frac{\pi(b-a)}{\beta}\right)\right) & T<a, \\
\frac{c}{6}\log\left(\frac{\beta^3}{2\pi^3\delta^2\epsilon}\left\{\sinh\frac{2\pi}{\beta}(b-a) - \sinh\frac{2\pi}{\beta}(b-T) - \sinh\frac{2\pi}{\beta}(T-a)\right\}\right) &\hspace{-0.2cm} a<T<b, \\
\frac{c}{3}\log\left(\frac{\beta}{\pi \delta}\sinh\left(\frac{\pi(b-a)}{\beta}\right)\right) & T>b,
\end{cases}
\end{equation}
while the contribution from disconnected geodesics is
\begin{equation}
S_{\text{disc}} =
\begin{cases}
\frac{c}{6}\log\left(\frac{\beta^2}{\pi^2\delta^2}\sinh\frac{2\pi a}{\beta} \sinh\frac{2\pi b}{\beta}\right) & T<a,\\
\frac{c}{6}\log\left(\frac{2\beta^3}{\pi^3\delta^2\epsilon}\sinh\frac{2\pi b}{\beta}\sinh\frac{\pi}{\beta}(T-a)\sinh\frac{\pi}{\beta}(a+T)\right) &\hspace{-0.5cm} a<T<b,\\
\frac{c}{6}\log\left(\frac{4\beta^4}{\pi^3\delta^2\epsilon^2}\sinh\frac{\pi}{\beta}(a-T)\sinh\frac{\pi}{\beta}(b-T)\sinh\frac{\pi}{\beta}(a+T)\sinh\frac{\pi}{\beta}(b+T)\right) & T>b,
\end{cases}
\end{equation}
which in the zero temperature limit $\beta\rightarrow\infty$ reproduces the results for ground state joining quenches \cite{Ugajin:2013xxa, Shimaji:2018czt}:
\begin{equation}
S_{\text{con}} = 
\begin{cases}
\frac{c}{3}\log\left(\frac{(b-a)}{\delta}\right) & T<a, \\
\frac{c}{6}\log\left(\frac{2(b-a)(T-a)(b-T)}{\delta^2\epsilon}\right) & a<T<b, \\
\frac{c}{3}\log\left(\frac{(b-a)}{ \delta}\right) & T>b,
\end{cases}
\end{equation}
\begin{equation}
S_{\text{disc}} = 
\begin{cases}
\frac{c}{6}\log\left(\frac{4a b}{\delta^2}\right) & T<a, \\
\frac{c}{6}\log\left(\frac{4b(T^2-a^2)}{\delta^2\epsilon}\right) & a<T<b, \\
\frac{c}{6}\log\left(\frac{4(T^2-b^2)(T^2-b^2)}{ \delta^2\epsilon^2}\right) & T>b.
\end{cases}
\end{equation}
Depending on the position of the entangling region with respect to $x=0$, the early time behavior can be dominated either by the connected or disconnected contributions, however at late times the connected contributions is always preferred.
\begin{figure}[H]
\centering
  \includegraphics[scale=0.5]{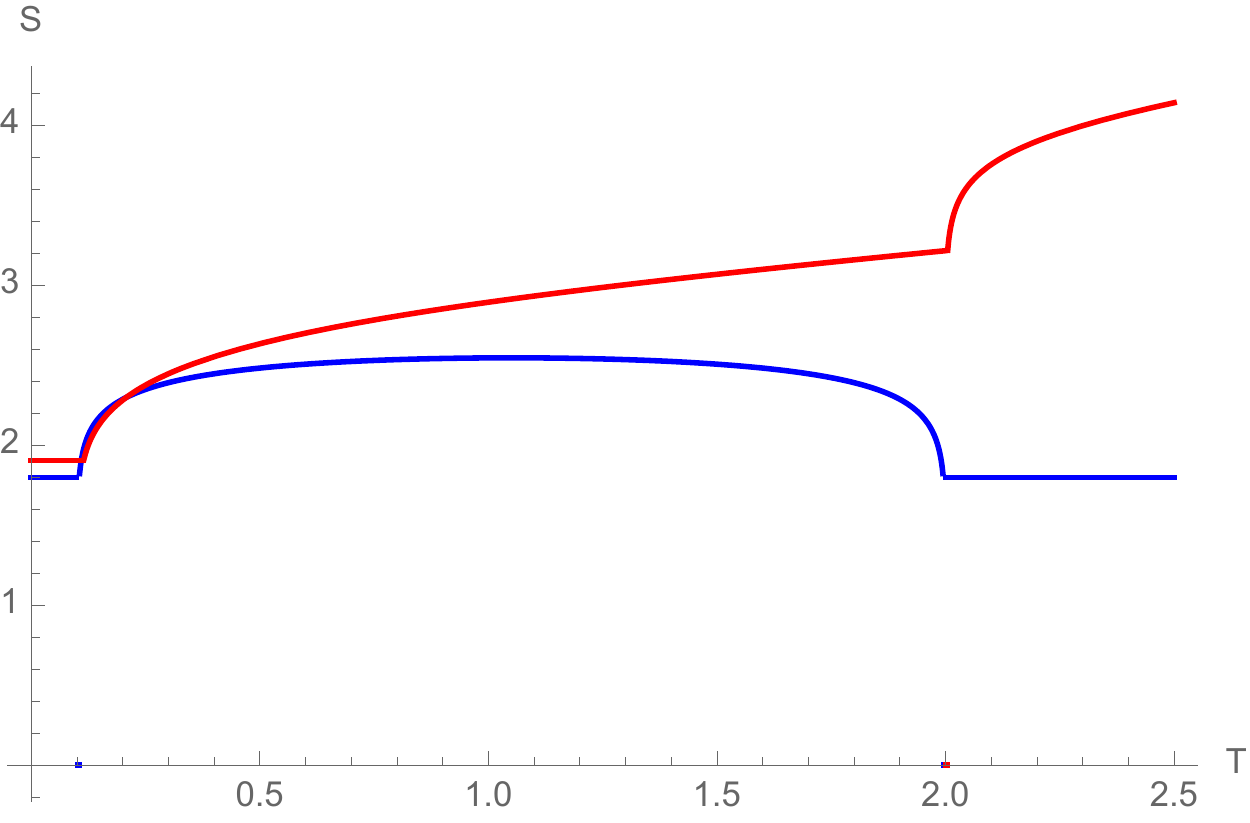}~~
  \includegraphics[scale=0.5]{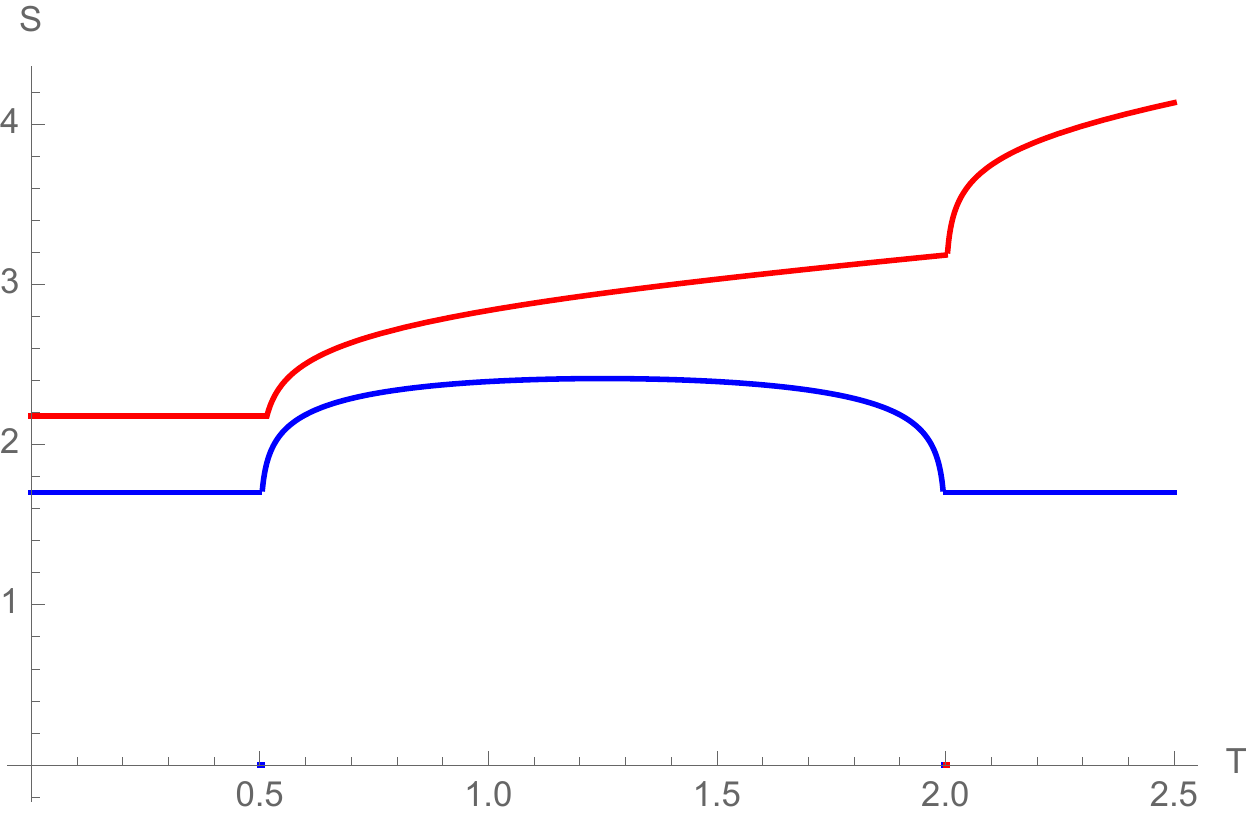}
  \caption{Connected (Blue) and Disconnected (Red) contributions to the entanglement entropy for $a=0.1$ (Left), $a=0.5$ (Right) and $b=2$}
\end{figure}

A similar calculation can be done for  $a<0$:
\begin{equation}
S_{\text{con}} = 
\begin{cases}
\frac{c}{6}\log\left(\frac{4\beta^4}{\pi^4\delta^2\epsilon^2}\sinh\frac{\pi}{\beta}(a-T)\sinh\frac{\pi}{\beta}(b-T)\sinh\frac{\pi}{\beta}(a+T)\sinh\frac{\pi}{\beta}(b+T)\right) & T<|a|, \\
\frac{c}{6}\log\left(\frac{2\beta^3}{\pi^3\delta^2\epsilon}\sinh\frac{\pi}{\beta}(T-a)\sinh\frac{\pi}{\beta}(b-T)\sinh\frac{\pi}{\beta}(b-a)\right) &\hspace{-0.3cm} |a|<T<b, \\
\frac{c}{3}\log\left(\frac{\beta}{\pi\delta}\sinh\frac{\pi}{\beta}(b-a)\right) & T>b,
\end{cases}
\end{equation}
\begin{equation}
S_{\text{disc}} = 
\begin{cases}
\frac{c}{6}\log\left(\frac{\beta^2}{\pi^2\delta^2}\sinh\frac{2\pi}{\beta}|a|\sinh\frac{2\pi}{\beta}b\right) & T<|a|, \\
\frac{c}{6}\log\left(\frac{2\beta^3}{\pi^3\delta^2\epsilon}\sinh\frac{2\pi}{\beta}b\sinh\frac{\pi}{\beta}(T-a)\sinh\frac{\pi}{\beta}(a+T)\right) &\hspace{-0.3cm} |a|<T<b, \\
\frac{c}{6}\log\left(\frac{4\beta^4}{\pi^4\delta^2\epsilon^2}\sinh\frac{\pi}{\beta}(T-a)\sinh\frac{\pi}{\beta}(T-b)\sinh\frac{\pi}{\beta}(T+a)\sinh\frac{\pi}{\beta}(b+T)\right) & T>b.
\end{cases}
\end{equation}

However, neither of these cases are suited for study in the scaling limit because of their short thermalization time, instead we consider a new entangling region, given by the semi-infinite interval $[b,\infty)$. For this case the only contribution comes from the disconnected surface:
\begin{equation}
S_{disc} =
\begin{cases}
\frac{c}{6}\log\left(\frac{\beta}{\pi\delta}\sinh\frac{2\pi}{\beta}b\right) & T<b,\\
\frac{c}{6}\log\left(\frac{\beta^2}{\pi^2\delta\epsilon}\left(\cosh\frac{2\pi}{\beta}T - \cosh\frac{2\pi}{\beta}b\right)\right) & T>b,
\end{cases}
\end{equation}
we can then take the scaling limit by again performing a rescaling $b\rightarrow\Lambda b$ and $T\rightarrow \Lambda T$:
\begin{equation}
\label{JoningMembrane}
S_{disc} =
\begin{cases}
\frac{c\pi}{3\beta}b + \mathcal{O}(\log\beta) & T<b,\\
\frac{c\pi}{3\beta}T + \mathcal{O}(\log\beta) & T>b,
\end{cases}
\end{equation}
where we can recognize the prefactor $\frac{c\pi}{3\beta}$ as the entropy density for a $\text{CFT}_2$ in the high temperature limit, as given by the Cardy formula \cite{Iqbal:2016qyz, PhysRevLett.56.742, Hartman:2014oaa}. For $d=2$, the membrane tension is degenerate, $\mathcal{E}(v)=1$ and $v_B=1$, hence \eqref{JoningMembrane} exactly agrees with \eqref{EntTime} computed for a joining quench whose details are somewhat different from the protocol implemented here. This result hints at universality of joining quenches in the scaling limit. Next, we attempt to derive a membrane description of this process starting from the bulk dual geometry, and will find that it differs from the membrane description of the simplest joining quench protocol summarized in Fig.~\ref{fig:eow}.

\subsubsection{Bulk geometry for the thermal joining quench}
In the previous section we computed the holographic entanglement entropy in a section of $AdS_3$ spacetime and then apply the map $w\rightarrow f(w)$ to the final result. We can also consider what is the holographic dual of the space before the conformal maping into the half-plane. 

We begin with $AdS_3$ in infalling coordinates:
\begin{equation}
ds^2 = \frac{1}{Z^2}\left(-dV^2 - 2dVdZ + dX^2\right).
\end{equation}
We then apply a large diffeomorphism that extends the conformal map \eqref{ConfMap} into the bulk and that gives a metric in  infalling gauge studied in \cite{Compere:2015knw,Sheikh-Jabbari:2016unm}:
\begin{equation}
\label{BanadosMetric}
ds^2 = \frac{1}{z^2}\left[-\left(1-2M(t,x) z^2\right)dt^2 -2dvdz + 2J(t,x)dtdx+dx^2\right],
\end{equation}
with $M=L_+ + L_-$, $J=L_+ - L_-$ and
\begin{equation}
L_{\pm} = \frac{3 f(x\pm t)''^2-2f(x\pm t)'f(x\pm t)'''}{4f(x\pm t)'^2}.
\end{equation}
We found the appropriate diffeomorphism by working perturbatively in $z$. We followed a similar computation performed by \cite{Roberts:2012aq}, who worked in  Fefferman-Graham gauge. The result is:
\begin{equation}
\label{bulkMap}
\scriptsize
\begin{aligned}
V(x,t,z) &= \frac{1}{2}\left(f(x+t) - f(x-t)\right) + \frac{z f'(x-t)f'(x+t)\left(f'(x-t)+f'(x+t)-2\sqrt{f'(x-t)f'(x+t)}\right)}{zf'(x+t)f''(x-t)+f'(x-t)\left(2f'(x+t)-z f''(x+t)\right)},\\
X(x,t,z) &= \frac{1}{2}\left(f(x+t) + f(x-t)\right)+ \frac{z f'(x-t)f'(x+t)\left(-f'(x-t)+f'(x+t)\right)}{zf'(x+t)f''(x-t)+f'(x-t)\left(2f'(x+t)-z f''(x+t)\right)},\\
Z(x,t,z) &= \frac{2z\left(f'(x-t)f'(x+t)\right)^{3/2}}{zf'(x+t)f''(x-t)+f'(x-t)\left(2f'(x+t)-zf''(x+t)\right)}.
\end{aligned}
\end{equation}

The condition $X > 0$ is translated, using \eqref{bulkMap}, into a non-trivial condition for $z$. The spacetime then has a boundary (an end of the world brane) located at $z_\text{EOW}(x,t)$, given by solving $X(x,t,z_\text{EOW})=0$.
Furthermore, in $AdS_{3}$ the connected geodesics follow semi-circular trajectories which in infalling coordinates can be parametrized as:
\begin{equation}
\begin{aligned}
X_\text{HRT}(\lambda) &= X(b,T)(1-\lambda),\\
V_\text{HRT}(\lambda) &= V(b,T) - X(b,T)\sqrt{\lambda(2-\lambda)},\\
Z_\text{HRT}(\lambda) &= X(b,T)\sqrt{\lambda(2-\lambda)},
\end{aligned}
\end{equation}
with $0\leq\lambda\leq 1$, the geodesic begins on the conformal boundary for $\lambda=0$ and ends on the end-of-the-world brane at $\lambda=1$. One can then numerically invert the map \eqref{bulkMap} along this trajectory to obtain the geodesic on the original spacetime \eqref{BanadosMetric}.

To obtain a membrane theory description, one needs to carefully analyze geodesics to learn what portions of them are important, and then to implement the appropriate scaling on these portions \cite{Mezei:2016zxg,Mezei:2018jco}. The qualitative behavior of the geodesics is different for the cases $T<b$ and $T>b$ as can be seen in the following figures.
\begin{figure}[H]
\centering
  \includegraphics[scale=0.52]{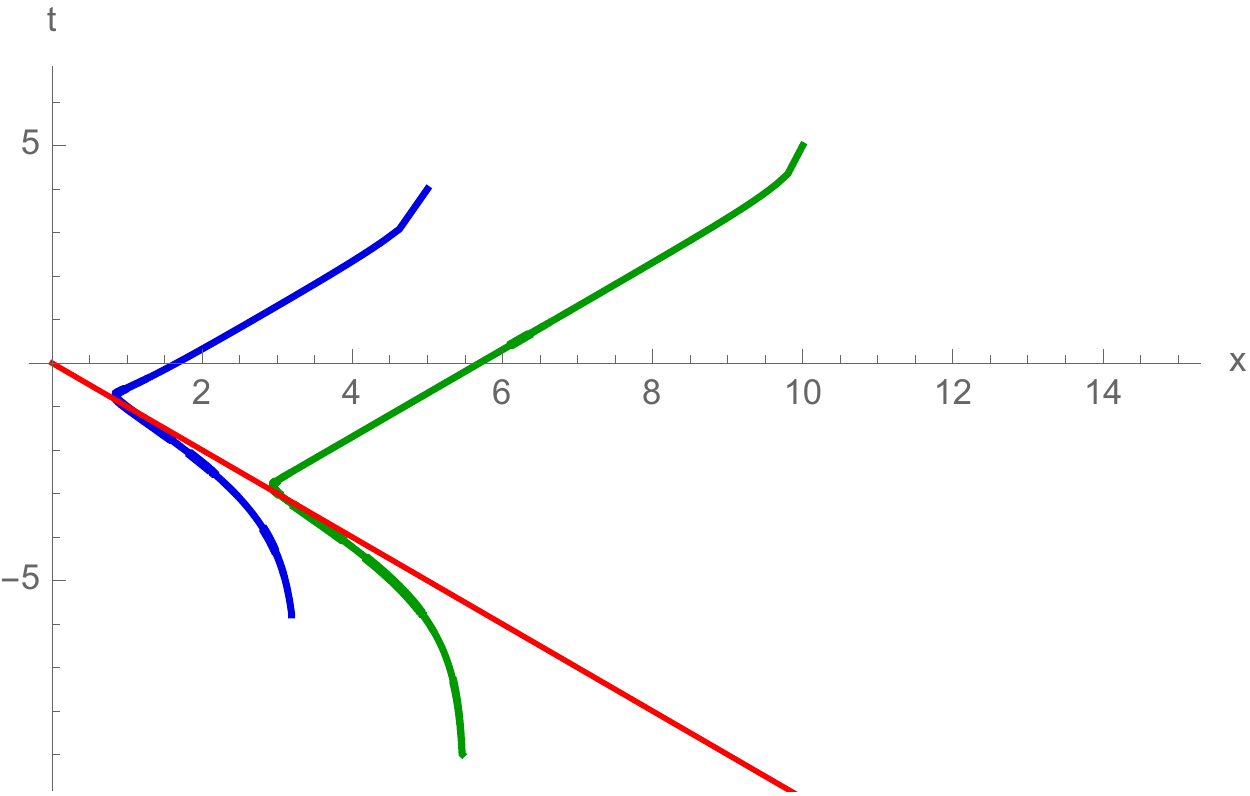}
   \includegraphics[scale=0.52]{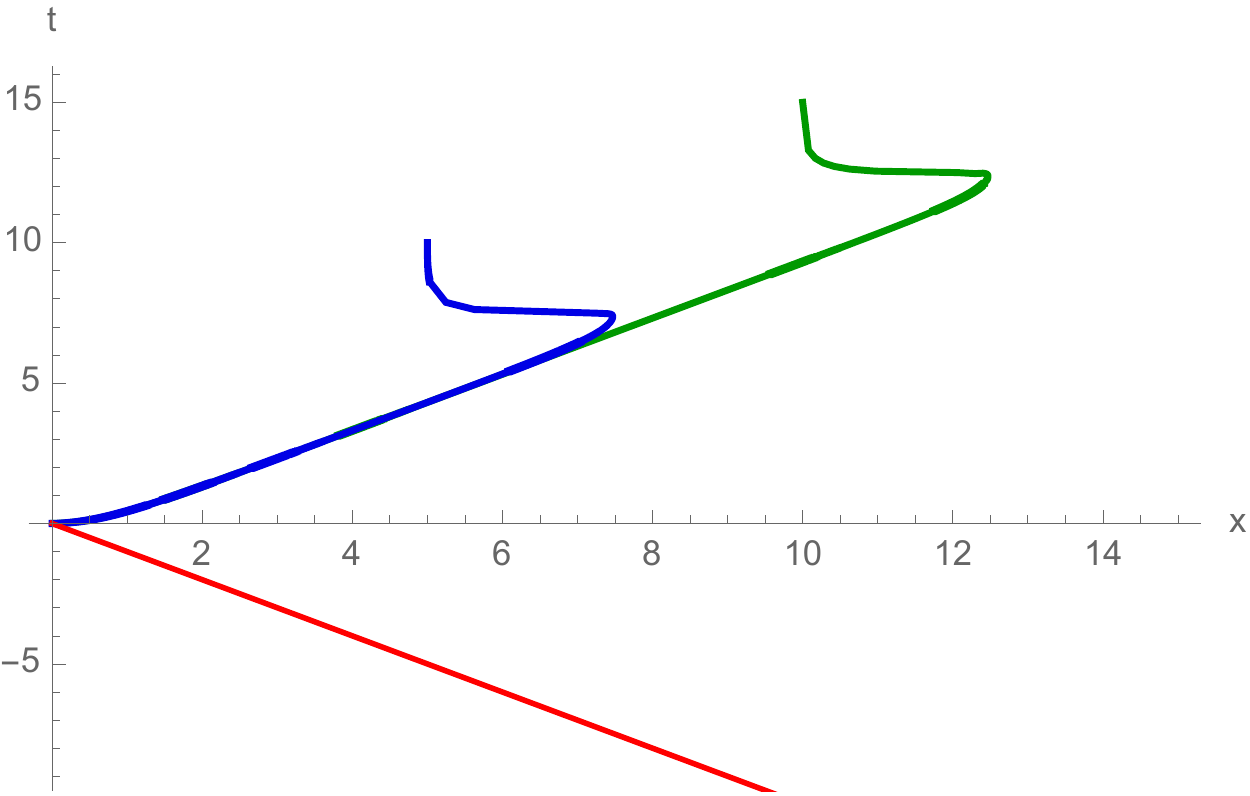}
  \caption{Geodesics for $(b,T) = (5,4),(10,5)$ (left) and $(5,10),(10,15)$ (Right). In red we show the end of the world brane. Here we consider a projection in the $z$ coordinate, which we expect to become the membrane in the scaling limit.}
  \label{fig:2dJoining}
\end{figure}
\begin{figure}[H]
\centering
  \includegraphics[scale=0.7]{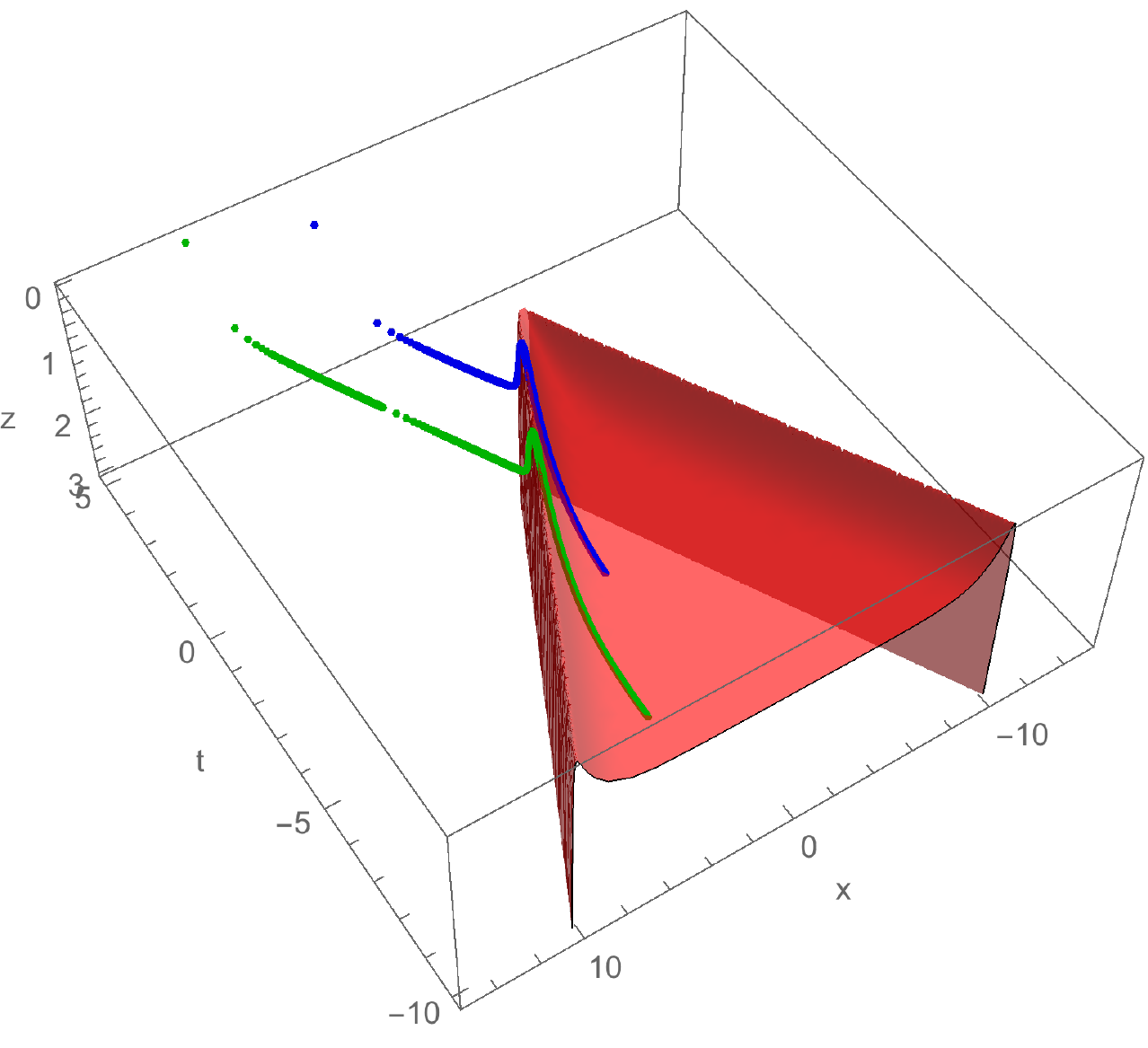}
  \caption{Geodesics for $(b,T) = (5,4),(10,5)$. Unlike in the projection in the $z$ coordinate, here we see that the geodesic does not intersect the end of the world brane but goes around it.}
  \label{fig:3dJoining}
\end{figure}
We see that for $T>b$ the behavior is very similar to the one in the previous model, with the geodesic ending on the tip of the end of the world brane, however we still see some interesting behavior, since the geodesic does not seem to go directly to the tip. The behavior for $T<b$ is even stranger, the geodesic goes around the end of the world brane and then approaches it asymptotically. In this later case the inversion of the map \eqref{bulkMap} becomes impossible for $\lam$ near $1$, since the geodesic seems to leave the patch cover by the coordinates $(z,t,x)$. Since we do not know how to incorporate all these features into the membrane theory, we do not describe more details of these geodesics further.

\section{Entanglement entropy of growing operators}\label{sec:OpGrowth}

Another interesting inhomogeneous setup is to consider the entanglement entropy of a time evolved local operator:
\es{Heisenberg}{
\sO(t,x)=e^{-iHt}\sO(0,x) e^{iHt}\,.
} 
Let us review the concept of operator entanglement. There is a one to one map between operators and states in the doubled Hilbert space $\sH=\sH_L\otimes \sH_R^*$:
\es{OpState}{
\sO \quad \leftrightarrow \quad \ket{\sO}=\sum_{i,j}\bra{i}\sO\ket{j}\, \ket{i,j}_{\sH}\,,
}
where the matrix element is taken in one copy of the system. What we mean by entanglement entropy of an operator is the entanglement entropy of the state $\ket{\sO(x,t)}$. A simple state to consider is the maximally entangled state $\ket{\mathbb{I}}=\sum_{i} \ket{i,i}_{\sH}$.

In QFT these notions requires regularization. The regularized maximally entangled state is the thermofield double state $\ket{\text{TFD}}\equiv\ket{e^{-\beta H/2}}=\sum_{n} e^{-\beta E_n/2}\ket{n,n}_{\sH}$. Local operators require smearing, which can be conveniently implemented by Euclidean evolution: 
\es{Heisenberg2}{
\sO_\beta(t,x)=\sO(t,x)e^{-\beta H/2}  \quad \leftrightarrow \quad \ket{\sO_\beta(t,x)}=\sO_L(t,x) \ket{\text{TFD}}\,,
} 
where we noticed that the operator has the interpretation of acting on the LHS of the TFD state.\footnote{There are other possible regularization prescriptions schemes, e.g. the ordering $e^{-\beta H/2}\sO(t,x)$ gives  $\sO_R(t,x) \ket{\text{TFD}}$. }${}^,$\footnote{The Heisenberg evolution is defined by $\sO(t)\equiv e^{iHt}\sO(0)e^{-iHt}$. } 

The gravitational dual of the TFD state is the eternal black hole \cite{Maldacena:2001kr}, while that of $\ket{\sO_\beta(t_\sO,0)}$ is the localized shock spacetime of \cite{Roberts:2014isa}, see \cite{Dray:1985yt,Schoutens:1993hu} for important early literature. To get the operator entanglement, we have to determine the extremal area HRT surfaces anchored on the boundary at $t=0$ on $A_{L} \cup A_R$. For $t_{\sO}\,, x\gg \beta$, the limit we are interested in, it is well approximated by the metric:
\es{ShockMetric}{
ds^2&=2A(uv)dudv+B(uv)dx^2-2A(uv)h(x)\delta(u)du^2\,, 
}
where $u,v$ are Kruskal coordinates, and 
\es{hexpr}{
h(x)\propto {1\ov G_N}\,{e^{{2\pi\ov \beta}\le(t_{\sO}- \abs{x}/v_B\ri)}\ov \abs{x/\beta}^{d-2\ov 2}}\,, \qquad v_B={2\pi\ov \beta}\,\sqrt{2A(0)\ov (d-1)B'(0)}\,.
}
That is we have two black half eternal black branes glued together along their horizon with the shift in $v$ equalling $h(x)$. To get to the membrane theory description, the {\it outgoing} Eddington-Finkelstein coordinate system is more appropriate:\footnote{For completeness we give explicit expressions for various quantities of interest: 
\es{Various}{
\beta={4\pi\ov \abs{a'(1)}}\,, \qquad v_B=\sqrt{ \abs{a'(1)}\ov 2(d-1)}\,.
}}
\es{Infalling}{
  u_{L,R}&=\pm e^{-{2\pi \ov \beta} t_{L,R}}\,, \qquad  uv=-e^{{4\pi \ov \beta} \,z_*(z)}\,,  \qquad z_*(z)\equiv\int^z {dz'\ov a(z') b(z')} \,,\\
  ds^2_{L,R}&={1\ov z^2}\le[-a(z)dt_{L,R}^2+{2\ov b(z)}\, dz dt_{L,R}+dx^2\ri]\,.
}
A key observation is that for the region of interest, $1< z \leq z_*$, we have $uv=O(1)$, hence in the scaling limit we have to have $\log u\approx -\log v$. This means that we can think of $v_{L,R}\approx \pm e^{ {2\pi \ov \beta} t_{L,R}}$. 

\begin{center}
\begin{figure}[!h]
\centering
\includegraphics[scale=0.8]{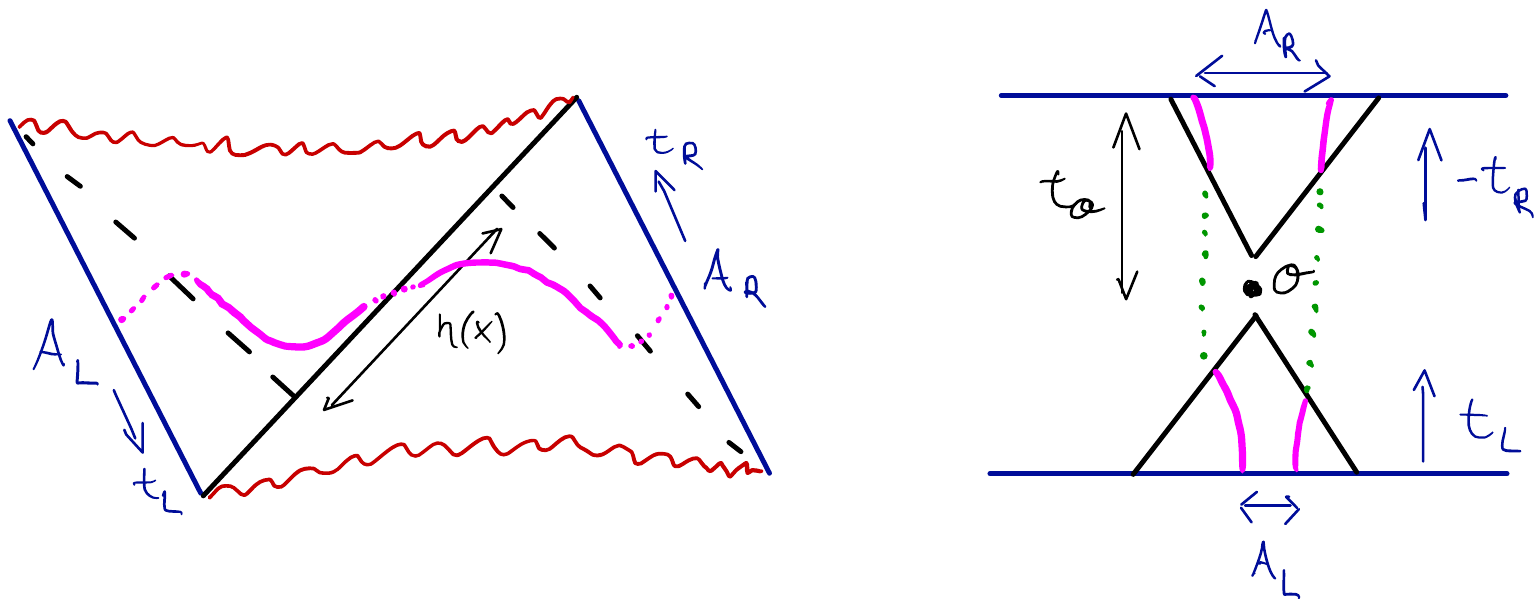}
\caption{{\bf Left:} The Penrose diagram of the shockwave spacetime is obtained by gluing together two black branes along their $u=0$ horizon (solid black line) with the shift in the space dependent shift in the  $v$ coordinates equalling $h(x)$. This is indicated by the mismatch of the $v_{L,R}$ horizons drawn by dashed black lines. With magenta we sketch the HRT surface interpolating between the regions $A_{L}$ and $A_R$, the solid portions of the line are indicating the parts of the HRT surface captured by the membrane theory, while the dotted portions are discarded in the effective theory. {\bf Right:} The membrane theory description of the entanglement entropy of $\sO(t_\sO,0)$. The solid portions of the HRT surfaces map onto the membranes living in the two cones, whose faces are glued together as indicated by the dotted green lines. The tips of the cones are displaced from $\sO$ by $t_\text{scr}$.
\label{fig:shockwave}}
\end{figure}
\end{center}

\begin{center}
\begin{figure}[!h]
\centering
\includegraphics[scale=0.4]{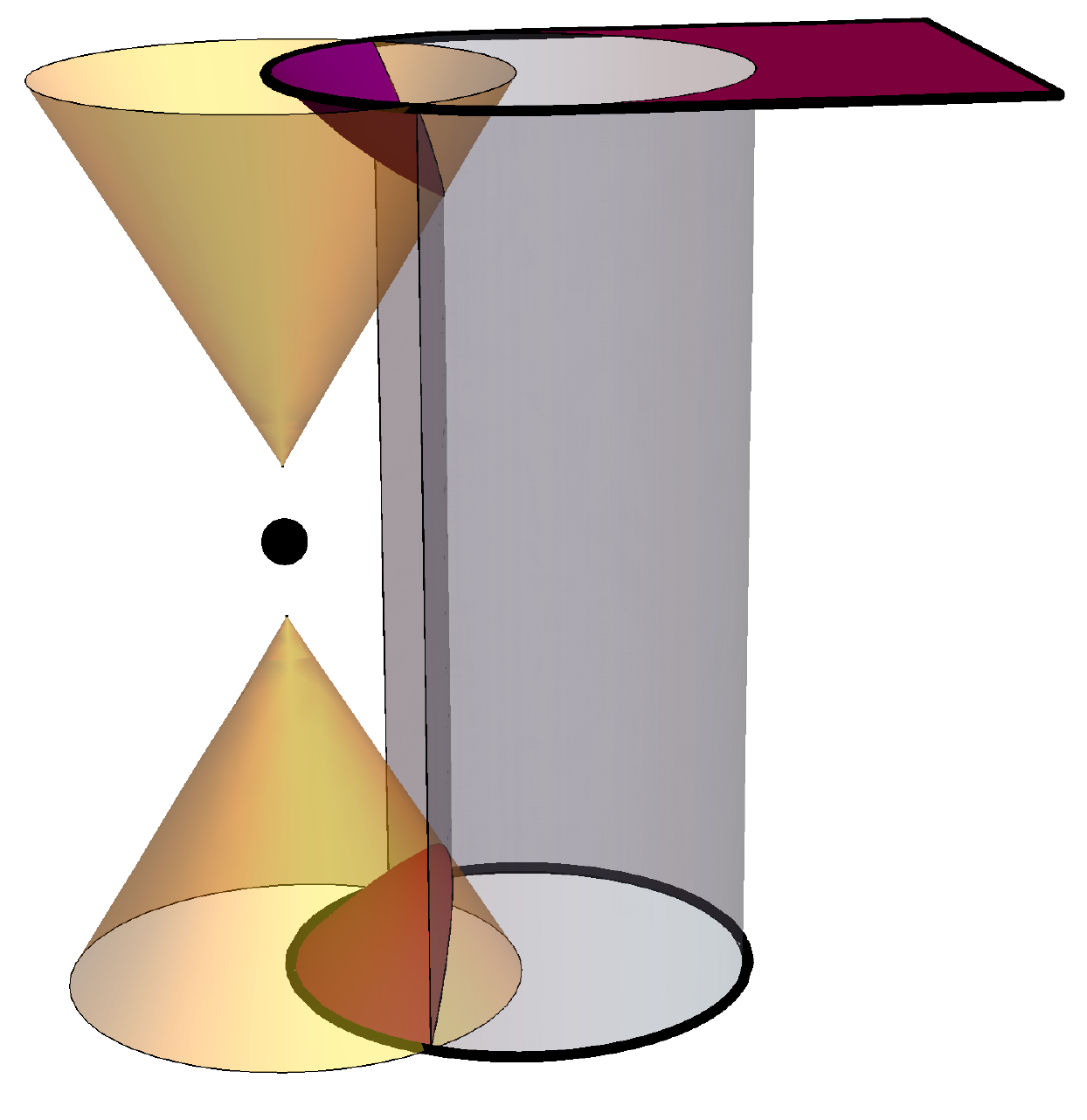}
\caption{ A three-dimensional version of the right figure of Fig.~\ref{fig:shockwave}. $A_L$ is chosen to be a circle, while $A_R$ is of a keyhole shape. The purple parts of the membrane inside the yellow cones are glued together as indicated by the grey surface. Since $A_L$ and $A_R$ do not overlap, hence we have to include a horizontal section of the membrane drawn with purple that also contributes to the entropy.
\label{fig:timefold}}
\end{figure}
\end{center}

Now let us consider what the HRT surface is doing in this spacetime, see Fig.~\ref{fig:shockwave}. The HRT surface connects to the boundary regions $A_{L,R}$ by cylinder-like portions that are marked by dotted purple lines on the Penrose diagram. These portions only contribute an area worth of entropy, and just like in the familiar quench setup, they are not captured by the membrane theory. The important parts of the HRT surface, drawn by solid purple line) is in the ``white hole'' region of the respective black branes (note that $t_L$ runs downwards). These can be parametrized by large values of the outgoing times $t_{L,R}$. These portions individually are identical to membranes in the familiar quench setup. The nontrivial physics comes from the way they are glued together. There is a shift $\Delta v\equiv v_L-v_R=h(x)$ between the left and right Kruskal coordinates across the $u=0$ horizon where the shockwave lies. To get a continuous HRT surface, it has to obey this matching condition, $v_L=v_R+h(x)$. This leaves the joining curve $v_R(\Omega)$ as an arbitrary timelike variable that we have to maximize over. Following the arguments in Appendix C of \cite{Roberts:2014isa}, as well similar maximization computations in \cite{Shenker:2013pqa,Stanford:2014jda}, we conclude that the result of this maximization is $ v_R\approx -v_L \approx h(x)/2$, i.e. the crossing point is halfway between the shifted $v_{L,R}=0$ horizons. Since  $v_{L,R}\approx \pm e^{{2\pi\ov \beta}  t_{L,R}}$, this translates into the condition 
\es{Matching}{
t_L\approx t_R\approx {\beta\ov 2\pi } \log h(x)=t_{\sO}-t_\text{scr}- \abs{x}/v_B\,,
}
 where we defined the scrambling time $t_\text{scr}={\beta\ov 2\pi}\, \log G_N$ following \cite{Shenker:2013pqa,Stanford:2014jda,Roberts:2014isa}. We conclude that the membrane lives in two cones defined by the contours \eqref{Matching}, whose  faces are glued together. See Figs.~\ref{fig:shockwave} and~\ref{fig:timefold} for illustration, where the identifications are shown by a dotted green line and a gray surface respectively.

This is the time fold geometry found to be the intrinsic geometry of the maximal Cauchy slice through the geometry in \cite{Roberts:2014isa}. The same geometry was introduced in the membrane theory from the random circuit perspective in \cite{Jonay:2018yei}. The difference here is the presence of the scrambling time $t_\text{scr}$. It is remarkable that a growing local operator has a simple entanglement structure that can be captured by the membrane theory in a nontrivial glued cone or time fold geometry. Within the cones we have to work with the same membrane tension function ${\cal E}(v)$ as in other setups, there is no sensitivity to the operator $\sO$ and no substructure within the butterfly cone. Probing growing operators with out of time order correlation functions \cite{Shenker:2013pqa,Roberts:2014isa,Shenker:2014cwa}, one finds richer, but less universal structure within the butterfly cone \cite{Xu:2018xfz,Khemani:2018sdn,Mezei:2019dfv}. In view of this complicated physics, we find it remarkable that operator entanglement is so universal and simple.

\section{Higher derivative corrections}\label{sec:HD}
So far we have showed that the membrane theory is an accurate description for a large family of quench protocols. However, our analysis so far based on the holographic dual being Einstein gravity. In this section we show that the membrane theory can be generalized to also take into account more general theories of gravity and the corresponding generalization of the holographic entanglement entropy prescription. In the context of AdS/CFT, this can be seen as ${\alpha'}$ corrections to the original formula for entanglement entropy, corresponding to finite 't Hooft coupling.

We will consider gravity theories characterized by an action of the form:
\begin{equation}
I = \frac{-1}{16\pi G_N}\int d^{d+1}x\sqrt{g}\left[\left(R+\frac{d(d-1)}{ l^2}\right)+\lambda_1 R^2 + \lambda_2 R_{\mu\nu} R^{\mu\nu} + \lambda_3 R_{\mu\nu\rho\sigma}R^{\mu\nu\rho\sigma}\right],
\end{equation}
which lead to equations of motion that include up to forth order derivatives of the metric. An special case of this class of theories, whose equations of motion include only second order derivatives, is Gauss-Bonnet gravity \cite{doi:10.1063/1.1665613}:
\begin{equation}
I_{GB} =\frac{-1}{16\pi G_N}\int d^{d+1}x\sqrt{g}\left[\left(R+\frac{d(d-1)}{ l^2}\right)+ \lambda_{GB}\left( R^2 -4 R_{\mu\nu} R^{\mu\nu} +  R_{\mu\nu\rho\sigma}R^{\mu\nu\rho\sigma}\right)\right].
\end{equation}

For the following analysis it will be convenient to introduce a new coupling:
\begin{equation}
\Lambda_{GB} = \frac{(d-3)(d-2)\lambda_{GB}}{l^2},
\end{equation}
that is dimensionless and emphasizes the fact that, for $d\leq3$, the Gauss-Bonnet term does not modify the equations of motion, since it is a topological invariant. We also introduce:
\begin{equation}
\begin{aligned}
\Lambda_1 &= \frac{\lambda_1}{l^2} -\frac{\Lambda_{GB}}{(d-3)(d-2)},\\
\Lambda_2 &=\frac{\lambda_2}{l^2} + 4\frac{\Lambda_{GB}}{(d-3)(d-2)},\\
\end{aligned}
\end{equation}
in terms of which:
\begin{equation}
\begin{aligned}
I =-\frac{1}{16\pi G_N}\int d^{d+1}x\sqrt{g} &\vphantom{=} \left[(R+d(d-1))\right.\\
&\vphantom{=} \left.+ \frac{\Lambda_{GB}}{(d-3)(d-2)}\left( R^2 -4 R_{\mu\nu} R^{\mu\nu} +  R_{\mu\nu\rho\sigma}R^{\mu\nu\rho\sigma}\right) \right.\\
&\vphantom{=} \left. + \Lambda_1  R^2 + \Lambda_2  R_{\mu\nu}R^{\mu\nu}\right],
\end{aligned}
\end{equation}
 where we set the dimensionful parameter $l^2$ to one, we will work in this units from this point forward.
 
As before, we are interested on black brane solutions of the form:
\begin{equation}
ds^2 = \frac{l_{AdS}^2}{z^2}\left[-a(z)dt^2 - \frac{2}{b(z)}dtdz + dr^2 + r^2 d\Omega_{d-2}^2\right],
\end{equation}
where $a(z)$ and $b(z)$ are also functions of the higher curvature couplings. We impose the boundary conditions: $a(1)=0$ and $a(0)=b(0)=1$, in order to obtain an asymptotically anti-de Sitter black brane solution. \footnote{We notice that, even though we are now working in units of $l^2=1$, this does not fix the physical AdS radius $l_{AdS}$, since in general it will receive corrections from the higher derivative couplings.}

For the particular case of Gauss-Bonnet gravity, it is possible to find a charge neutral black brane solution, given by \cite{Cai:2001dz, Garraffo:2008hu}:
\begin{equation}
\label{GBsol}
\begin{aligned}
a(z) &=\frac{\left(1-\sqrt{1-4(1-z^d)\Lambda_{GB}}\right)}{\left(1-\sqrt{1-4\Lambda_{GB}}\right)},\\
b(z) &= 1,\\
l_{AdS}^2 &= \frac{2\Lambda_{GB}}{1-\sqrt{1-4\Lambda_{GB}}}.
\end{aligned}
\end{equation}

For the more general case with $\Lambda_1, \Lambda_2 \neq 0$, no black brane solution is known, however we can consider a perturbative solution which, to second order in the higher derivative couplings, takes the form:
\begin{equation}
\label{gravsol}
\begin{aligned}
a(z) &=(1-z^d)\left[1-z^d \Lambda_{GB}\left(1+(3-2z^d)\Lambda_{GB}+2d^2\left(\frac{(d^2-5)}{2d(d-1)}-z^d\right)\Lambda_1 \right.\right. \\ &- \left.\left. \frac{d}{d-1}(2d^2-5d+5)\Lambda_2\right)\right],\\
b(z) &= 1+2d z^{2d}\Lambda_{GB}\left((2d+1)\Lambda_1 + (d+1)\Lambda_2\right),\\
l_{AdS} &= 1- \frac{1}{2}\Lambda_{GB} - \frac{d(d-3)}{2(d-1)}\left((d+1)\Lambda_1+\Lambda_2\right) \\ &- \frac{5}{8}\left(\frac{d(d-3)}{d-1}\left((d+1)\Lambda_1+\Lambda_2\right)+\Lambda_{GB}\right)^2,
\end{aligned}
\end{equation}
where again we see that no correction occurs if $d\leq3$. It is easy to see that, for $\Lambda_1 = \Lambda_2 = 0$, this agrees with the expansion of the full solution \eqref{GBsol}. We notice that corrections proportional to $\Lambda_1$ and $\Lambda_2$ appear only to second order and contribute only if $\Lambda_{GB}\neq0$.

\subsection{Entanglement with higher derivatives in the scaling limit}
With the black brane solutions in hand, we can compute the entanglement entropy in the same way as described in Sec.~\ref{sec:Review}, being careful to notice that for gravitational theories with higher derivatives the original prescription for the area functional is modified to \cite{Camps:2013zua,Dong:2013qoa}:
\begin{equation}
\begin{aligned}
S &= \frac{1}{4 G_N}\int d^{d-1}y \sqrt{\vert \gamma\vert}\\
&\times\left(1+2\lambda_1 R + \lambda_2\left(R^a_a-\frac{1}{2}K^aK_a\right) + 2\lambda_3\left(R^{ab}_{ab} - K^{a\mu\nu}K_{a\mu\nu}\right)\right),
\end{aligned}
\end{equation}
where the codimension two surface is characterized by two normal vectors $n_a^{\mu}$, $R_{abcd} = n^{\mu}_a n^{\nu}_b n^{\rho}_c n^{\sigma} _d R_{\mu\nu\rho\sigma}$, and $K^{a}_{\mu\nu}$ is the extrinsic curvature. We evaluate this functional in the scaling limit by performing the rescaling \eqref{CoordScaling}.

We notice an important simplification:  the Gauss-Bonnet term is always subleading in $1/\Lambda$, hence:
\begin{equation}
S = \frac{\Lambda^{d-1}}{4 G_N}\int d^{d-1}y \sqrt{\vert \gamma\vert}\left(1+2\Lambda_1 R + \Lambda_2\left(R^a_a-\frac{1}{2}K^aK_a\right)\right),
\end{equation}
or more explicitly:
\begin{equation}
\label{functional}
\begin{aligned}
S &= \frac{\Lambda^{d-1}}{4G_N}\int d\tau d\Omega \frac{r^{d-2}}{z^{d-1}}\sqrt{\left(1+\frac{(\partial_{\Omega}r)^2}{r^2}\right)\left(v^2-a(z)\right)}\\
&\times\left(1 + 2\Lambda_1 F_1(z,v) + \Lambda_2 F_2(z,v)\right),\\
F_1(z) &=-b(z ) \left(z  \left(z  b(z ) a''(z )+a'(z ) \left(z 
   b'(z )-2 d b(z )\right)\right) \right. \\ 
 &+  \left. a(z ) \left(d (d+1) b(z )-2 d z b'(z )\right)\right),\\
F_2(z,v) &= \frac{b(z)}{8(v^2-a(z))^2}\left(H_0(z)+v^2 H_1(z)+v^4H_2(z)\right),\\
H_0(z) &=a(z ) \left(-z ^2 b(z ) a'(z )^2-4 z  a(z ) \left(z  b(z
   ) a''(z )+a'(z ) \left(z  b'(z )-2 (d+1) b(z )\right)\right)\right.\\
   &-\left.4
   (d+1) a(z )^2 \left((d+1) b(z )-2 z  b'(z )\right)\right),\\
  H_1(z) &= 4 a(z ) \left(z  \left(3 z  b(z ) a''(z )+a'(z ) \left(3 z 
   b'(z )-4 (d+1) b(z )\right)\right)\right.\\
   &+\left. 2 (d+1) a(z ) \left((d+1) b(z )-2
   z  b'(z )\right)\right),\\
   H_2(z) &= 8 z  \left(a'(z ) \left((d+1) b(z )-z  b'(z )\right)-z  b(z
   ) a''(z )\right)\\
   &-4 (d+1) a(z ) \left((d+1) b(z )-2 z  b'(z
   )\right).
\end{aligned}
\end{equation}

Let us consider first the simplest case of $\Lambda_1=\Lambda_2=0$, for which all corrections come from the change in the metric. The algebraic equation of motion for $z$ is then the same as in the case without higher derivatives:
\begin{equation}
v^2 =c(z)= a(z) - \frac{z a'(z)}{2(d-1)}.
\end{equation}
In terms of which:
\begin{equation}
\begin{aligned}
S&= s_{th}\Lambda^{d-1}\int d\tau d\Omega r^{d-2}\sqrt{\left(1+\frac{(\partial_{\Omega}r)^2}{r^2}\right)}\mathcal{E}(v),\\
&=s_{th}\Lambda^{d-1}\int d^{d-1}y\sqrt{-\ga}\, \frac{\mathcal{E}(v^2)}{\sqrt{1-v^2}},\\
\mathcal{E}(v) &= \left.\sqrt{\frac{-a'(z)}{2(d-1)z^{2d-3}}}\right|_{z=c^{-1}(v^2)}.
\end{aligned}
\end{equation}

From this expression we can obtain the entanglement velocity by evaluating $\mathcal{E}(0)=v_E$  \cite{Mezei:2018jco}. This requires solving the equation $c(z_*)=0$, the full non-perturbative expression is:
\begin{equation}
\label{vEGB}
\begin{aligned}
v_E^2 &= \frac{d\Lambda_{GB}\left(\frac{(d-1)\left(2-d+4(3d-4)\Lambda_{GB}+\sqrt{(d-2)^2+4d(3d-4)\Lambda_{GB}}\right)}{(3d-4)^2\Lambda_{GB}}\right)^{\frac{2-d}{d}}}{(d-1)\left(1-\sqrt{1-4\Lambda_{GB}}\right)} \\ &\times \frac{1}{\sqrt{(1-4\Lambda_{GB})+\frac{4(d-1)\left(2-d+4(3d-4)\Lambda_{GB}\right)+\sqrt{(d-2)^2+4d(3d-4)\Lambda_{GB}}}{(3d-4)^2}}}.
\end{aligned}
\end{equation}

As it stands this expression is not very insightful, so it is useful to consider its expansion for small $\Lambda_{GB}$:
\begin{equation}
\label{GBvEexp}
\begin{aligned}
v_E &= v_E^{(0)}\left(1- \left(\frac{d-1}{d-2}\right)\Lambda_{GB}\right),\\
v_E^{(0)} &= \frac{\left(\frac{d-2}{d}\right)^{\frac{d-2}{2d}}}{\left(\frac{2(d-1)}{d}\right)^{\frac{d-1}{d}}},
\end{aligned}
\end{equation}
where $v_E^{(0)}$ is the entanglement velocity for the Schwarzschild solution \cite{Mezei:2018jco}.

Similarly, we can compute the butterfly velocity from either $\mathcal{E}(v_B) = v_B$ or $\mathcal{E}'(v_B)=1$, order by order in $\Lambda_{GB}$ \cite{Mezei:2018jco}. We can obtain the butterfly velocity nonperturbatively, by noting that in the membrane theory, $v_B$ is the largest possible value of $v$ in the physically relevant interval $1<z<z_*$. Since $c(z)$ is a monotonically decreasing function, then $v_B^2= c(1)$ :
\begin{equation}
\label{vBGB}
\begin{aligned}
v_B &= v_B^{(0)}\sqrt{\frac{1+\sqrt{1-4\Lambda_{GB}}}{2}},\\
v_B^{(0)} &= \sqrt{\frac{d}{2(d-1)}},
\end{aligned}
\end{equation}
where once again the prefactor $v_B^{(0)}$ corresponds to the butterfly velocity for the case of a black brane without higher derivatives. 

We remark once again that this is an exact result, and it is in agreement with previous calculations in the literature \cite{Roberts:2014isa}. It is straightforward to confirm that this velocity satisfies the equations $\mathcal{E}(v_B) = v_B$ and $\mathcal{E}'(v_B)=1$.

The next simplest case is given by  $\Lambda_1\neq0$ but $\Lambda_2=0$,  for which the functional is modified but the algebraic equation of motion for $z$ still contains only single powers of $v^2$. Because of this full analytic results are still available. 

From \eqref{functional}, the algebraic equation for $z$ is:
\begin{equation}
v^2 = c(z) = a(z) - \frac{z(1+2\Lambda_1 F_1(z))a'(z)}{2(d-1)(1+2\Lambda_1 F_1(z))-4\Lambda_1 z F_1'(z)}.
\end{equation}

Using this expression we can write the entanglement functional as: \footnote{It is important to notice that, in the definition of the entropy density $s_{th}$, we use not the Bekenstein formula but its generalization, given by the Wald formula, which takes into consideration corrections from higher derivatives \cite{Wald:1993nt}}
\begin{equation}
\begin{aligned}
S&=s_{th}R^{d-1}\int d^{d-1}y\sqrt{-\ga}\, \frac{\mathcal{E}(v)}{\sqrt{1-v^2}}\\
\mathcal{E}(v) &= \sqrt{\frac{-z^{3-2d}(1+2\Lambda_1 F_1(z))a'(z)}{2(d-1)(1+2\Lambda_1 F_1(z))-4z \Lambda_1 F_1'(z)}}\\
&\times\left. \left(\frac{1+2\Lambda_1 F_1(z)}{1+2\Lambda_1 F_1(1)}\right)\right|_{z-c^{-1}(v^2)}.
\end{aligned}
\end{equation}

We can then compute the butterfly velocity as before, evaluating the function $c(z)$ at $z=1$:
\begin{equation}
\label{vB1}
\begin{aligned}
v_B^2 &=  - \frac{(1+2\Lambda_1 F_1(1))a'(1)}{2(d-1)(1+2\Lambda_1 F_1(1))-4\Lambda_1  F_1'(1)}.
 \end{aligned}
\end{equation}

For general functions $a(z)$ and $b(z)$, expanded around $z=1$ and satisfying the appropriate boundary conditions, the expression \eqref{vB1} agrees with the previous result obtained in \cite{Mezei:2016wfz}. For our particular solution \eqref{gravsol}, we can obtain a result to next-to-leading order in the couplings:

\begin{equation}
\label{vB1per}
v_B = v_B^{(0)}\left(1 - \frac{\Lambda_{GB}}{2}-\frac{5}{8}\Lambda_{GB}^2 - \frac{d(d+1)(3d-5)}{2(d-1)}\Lambda_{GB}\Lambda_1\right).
\end{equation}

We notice that at leading order we have corrections only from the Gauss-Bonnet coupling, and the first two correction terms in the parenthesis agree with a perturbative expansion of \eqref{vBGB}.  Again we compute the entanglement velocity to next-to-leading order:
\begin{equation}
\label{vE1}
v_E = v_E^{(0)}\left(1 - \frac{d-1}{d-2}\Lambda_{GB} + \frac{(d ((5-2 d) d+3)-6) \Lambda _{GB}^2}{2 (d-2)^3}- \frac{d(d+1)(3d-5)}{d-2}\Lambda_1\Lambda_{GB}\right),
\end{equation}
where the first two correction terms are just the second order expansion of the full result for Gauss-Bonnet gravity \eqref{vEGB}.

With the experience obtained from the two previous cases, we can now turn to the general situation of higher derivative gravity with $\Lambda_1$ and $\Lambda_2$ non-zero. In this case one can in principle follow the same procedure as before, determining the equation of motion of $z$ from \eqref{functional}. However, in this case we do not have an equation of the form $v^2 = c(z)$, but an equation that depends on powers of $v^2$ up third order. This equation can still be solved in principle, since it is just a cubic equation on $v^2$, and from it one can still compute the entanglement and butterfly velocity and determine the energy function $\mathcal{E}(v)$. However, since the gravitational solutions themselves are perturbative, it will be easier to solve this problem order by order in the higher derivative couplings.

Before considering the perturbative expansion, we notice that a simplification occurs when evaluating $v_B^2$, since $H_0(1) = H_1(1) = 0$:

\begin{equation}
\label{GralvB}
\begin{aligned}
v_B^2 &= \frac{B_1(\Lambda_1,\Lambda_2)-\sqrt{B_1(\Lambda_1,\Lambda_2)^2-16b(1)\Lambda_2 H_0'(1)B_2(\Lambda_1,\Lambda_2)}}{4B_2(\Lambda_1,\Lambda_2)},\\
B_1(\Lambda_1,\Lambda_2)&= \left(8+16\Lambda_1 F_1(1)-3\Lambda_2 b(1)H_2(1)\right)a'(1)-2\Lambda_2 b(1) H_1'(1),\\
B_2(\Lambda_1,\Lambda_2)&=H_2(1) \Lambda _2 b'(1)-d \left(b(1) H_2(1) \Lambda _2+16 F_1(1) \Lambda
   _1+8\right)\\
   &+b(1) \Lambda _2 H_2'(1)+b(1) H_2(1) \Lambda _2+16 \Lambda _1
   \left(F_1'(1)+F_1(1)\right)+8.
\end{aligned}
\end{equation}

Which again agrees with previous calculations of $v_B$ for general metrics with appropriate boundary conditions \cite{Mezei:2016wfz}. 

In perturbation theory we can solve the cubic equation in $v^2$ to write it as:
\begin{equation}
\label{vPert}
\begin{aligned}
v^2 &= c(z) =  a(z) - \frac{z a'(z)}{2(d-1)}\\
&-\frac{d^2\Lambda_{GB}}{d-1}\left(2d z^d\Lambda_1+2\Lambda_2-2z^d\Lambda_2+dz^d\Lambda_2\right)z^{2d}.
\end{aligned}
\end{equation}

The first line is the known algebraic equation for the unmodified functional and it contains all order corrections from the metric. The second term is the correction from the change in the functional, it contains up to first order corrections from the metric and we notice it does not contribute to the algebraic equation to first order.This expression can then be evaluated at $z=1$ to compute the perturbative expansion of $v_B$, in agreement with our previous results:
\begin{equation}
\begin{aligned}
v_B^2 &= -\frac{a'(1)}{2(d-1)}-\frac{d^3}{(d-1)}(2\Lambda_1+\Lambda_2)\Lambda_{GB},
\end{aligned}
\end{equation}
where the first term is given by metric corrections to the unmodified functional and the second one comes from the modification of the functional.

Finally we can provide a form for the energy function as an implicit function of $v$:

\begin{equation}
\begin{aligned}
\mathcal{E}(v) &=\left. \sqrt{\frac{-a'(z)}{2(d-1)z^{2d-3}}}\left(1+\mathcal{E}_{(1)}(v)\right)\right|_{z = c^{-1}(v)},\\
\mathcal{E}_{(1)}(v) &=\ -d\Lambda_{GB}
\left(\left(2(2d-1)\Lambda_1+(d-2)\Lambda_2\right)z^{2d}+2d z^d\Lambda_2-2(d-1)(\Lambda_1+\Lambda_2\right).
\end{aligned}
\end{equation}
where the prefactor is the result coming from the unmodified action, containing all order corrections in the metric function, while the second term accounts for corrections on the functional itself, and so it contains only second order corrections. We can write this as an explicit function of $v$ to next-to-leading order in the higher curvature couplings:
\begin{equation}
\begin{aligned}
\mathcal{E}(v) &= \frac{v_E}{\left(1-v^2\right)^{\frac{d-2}{2d}}}\left(1 + \frac{2(d-1)^2}{d(d-2)}\Lambda_{GB}v^2\right.\\
&-\left.\frac{2 (d-1)^2 v^2 \left(d \left(d \left(5 d \left(v^2-1\right)-16 v^2+12\right)+9
   v^2\right)+2 v^2\right) \Lambda _{GB}^2}{(d-2)^3 d^2} \right.\\
   &+\left. \frac{2 (d-1) (3 d-5) v^2 \left(d \Lambda _1+\Lambda _1+\Lambda _2\right) \Lambda
   _{GB}}{d-2}\right),
\end{aligned}
\end{equation}
where
\begin{equation}
v_E = v_E^{(0)}\left(1 - \frac{d-1}{d-2}\Lambda_{GB}+\frac{(d ((5-2 d) d+3)-6) \Lambda _{GB}^2}{2 (d-2)^3}-\frac{d (3 d-5) \left(d \Lambda _1+\Lambda _1+\Lambda _2\right) \Lambda
   _{\text{GB}}}{d-2}\right),
\end{equation}
that we notice agrees with \eqref{vE1} when $\Lambda_2=0$.
\section{Subleading orders}\label{sec:NLO}

The membrane theory only captures the leading order extensive piece of the entropy in the large $R/\beta$ expansion. It is an interesting question, whether subleading orders in this expansion can be captured by the membrane theory.  Such corrections come from three sources: from the part of the HRT surface connecting the black brane horizon to the boundary, from subleading terms that come from the behind the horizon part of the static black brane patch of the geometry, and the part of the HRT surface that is in the genuinely time dependent part of the geometry and contributions from the initial state (represented by a pure AdS region in the Vaidya quench model or the near end of the world brane part of the geometry). In this section, we analyze the second source of corrections, and leave the others for future work. We find that the way the membrane theory captures these is analogous how higher derivative terms appear in the chiral Lagrangian or to higher gradient terms in hydrodynamics. The effects of some phenomenologically added higher derivatives terms were also considered in \cite{Jonay:2018yei}, here we derive their explicit form from holography.

Let us recall \eqref{AreaFunct}. In this section, after the rescaling \eqref{CoordScaling}, we want to keep the first subleading term in $\Lam$. 
\es{AreaFunctSub}{
S[A(T)]&={\Lam^{d-1}\ov 4G_N}\min_{m\sim A(T)}\int dt\, d\Omega \ {r^{d-2}\ov z^{d-1}}\sqrt{Q_0}\le[1+{1\ov \Lam}\,F(z,r)+\dots\ri]\\
Q_0(z,r)&\equiv \le[\dot{r}^2-\le(1+{(\pa_\Omega r)^2\ov r^2}\ri)a(z)\ri]\\
F(z,r)&\equiv{1\ov  Q_0(z,r)b(z)}\le[{\dot{r} \,(\pa_\Omega r \cdot \pa_\Omega z)\ov r^2}-\le(1+{(\pa_\Omega r)^2\ov r^2}\ri)\dot z\ri]\,.
}
As is familiar from perturbation theory, to obtain the first order correction to the on shell action, we can use the first order solution evaluated on the correction to the action. We obtain (after setting $\Lam=1$):
\es{AreaFunctSub2}{
S[A(T)]&=S_0[A(T)]+{1\ov 4G_N}\int dt\, d\Omega \ {\cal E}_0(v)\,F(z,r)\Big\vert_{z=c^{-1}(v^2)}+\dots\,.
}
Since $F(z,r)$ depends on the derivatives of $z(t,\Omega)$, it is clear that we cannot write this correction simply as a function of $v$, which is a function of $(n\cdot \hat{t})$. Instead, we have to consider derivatives of $n_\mu$, the extrinsic curvature tensor, $K_{\mu\nu}$. A term linear in $K_{\mu\nu}$ produces the right scaling $1/\Lam$; at higher orders we would also encounter the Riemann tensor, their powers and derivatives. Hence we expect that \eqref{AreaFunctSub2} can be written as $C_1(v) K^\mu_\mu+C_2(v) K_{\mu\nu} \hat{t}^\mu \hat{t}^\nu$.  Evaluating the expression explicitly, we only find the second structure:
\es{AreaFunctSub3}{
S[A(T)]&=S_0[A(T)]+{1\ov 4G_N}\int dt\, d\Omega \ {\cal E}_1(v)\, K_{\mu\nu}\hat{t}^\mu \hat{t}^\nu+\dots\,,\\
{\cal E}_1(v)&\equiv{(1-v^2)^{5/2}\ov {\cal E}_0(v)}\,\le[{1\ov b(z) \, z^{2(d-1)}}\, {dz\ov dv}\ri]\Bigg\vert_{c^{-1}(v^2)}<0\,.
}
It would be interesting to understand in what situations the extrinsic curvature term plays an important role in the physics of entanglement growth.  

\section{Summary, discussion and open questions}\label{sec:Summary}

In this paper we have significantly enlarged the domain of applicability of the membrane effective theory of entanglement dynamics. We studied quenches for inhomogeneous initial states: generically such states will also have inhomogeneities in conserved densities, whose dynamics is described by hydrodynamics. We derived how the membrane theory couples to hydrodynamics (but does not back react) in a beautiful geometric way. 

We also studied another inhomogeneous setup, the joining of two separately thermalized systems (living on a half space). 
Since the membrane theory Lagrangian is only sensitive to the conserved densities, and in the thermal joining quench hydrodynamics is trivial, we obtain nontrivial entanglement dynamics in this case from boundary conditions: there is a brane in the membrane theory Minkowski spacetime (descending from an end of the world brane in holography) on which the membrane can end. Our results precisely reproduce those of \cite{Jonay:2018yei} obtained from a transport equation reformulation of the problem, but complement it with a spacetime picture. We note that in the global quench it is also the boundary conditions that give the time dependence to the entanglement entropy (this is also partially true in inhomogeneous quenches), but there the brane is on the $t=0$ slice respecting the symmetry of the problem.

This joining quench setup makes it possible to address the recent claim that  black holes may not be fast scramblers based on entanglement, only based on out-of-time order correlators \cite{Shor:2018sws,Harrow:2019lyw}. In their setup, they consider a black hole cut in half and then joined, and based on a quantum circuit model of the photon sphere argue that entanglement saturation is slow. The spherical version of our model with an end of the world brane makes it clear that entanglement saturates in a time of order the light crossing time: this is the fastest possible entropy saturation time for a local quantum system.\footnote{A related end of the world brane model for a joining quench demonstrating fast saturation of the entropy was proposed independently by Juan Maldacena \cite{MaldacenaComment}.}

Another setup that we considered was an entanglement entropy of a time evolved local operator. There we found that the membrane spacetime is a double cone glued along their faces or equivalently a time fold. This same geometry was derived from  random circuits in  \cite{Jonay:2018yei}; our derivation applies to holographic gauge theories, and a distinguishing feature is that the scrambling time separates the tips of the cones, see Figs.~\ref{fig:shockwave} and~\ref{fig:timefold}. In this geometry, we have to use the same membrane tension, as in other setups. The same geometry was found to be that of the maximal spatial slice through the shockwave spacetime dual to the growing operator in \cite{Roberts:2014isa}, and it was related to the geometry of a minimal tensor network reproducing the operator.

The connection between tensor networks and the membrane theory is very direct: the membrane can be thought of as a coarse grained cut through the tensor network representing the state whose subsystem entropies we are computing. This is in fact how the membrane description is derived in the random circuit approach \cite{Nahum:2016muy,Jonay:2018yei}. However, the connection between the tensor network and the bulk geometry is more subtle, than envisioned in \cite{Roberts:2014isa}: the HRT surfaces (that become the membranes) do not lie in the same Cauchy slice, so it is not the geometry of the maximal volume slice that enters the membrane theory, but the geometry of the spacetime in the sliver between $z_h\leq z \leq z_*$ that gets reprocessed into an effective tensor network description. This geometry determines both the background spacetime and the angle dependent membrane tension ${\cal E}(v)$. It would be very interesting to better incorporate this way of thinking into the relation between tensor network approaches and holography \cite{Swingle:2009bg,Swingle:2012wq,Pastawski:2015qua,Hayden:2016cfa}.

An important direction in the exploration of the membrane theory is to enlarge the set of theories for which it can be derived as an effective description. To this end, we showed that finite coupling corrections to holographic gauge theories do not change the structure of the theory, by deriving the membrane tension function from the holographic entropy functional of higher derivative gravity theories. The entropy functional contains higher derivative terms, which may have been guessed to give rise to higher derivative terms in membrane theory. Instead,  we found that only the explicit form of ${\cal E}(v)$ and its relation to the geometry changed, but the membrane theory Lagrangian does not contain higher derivative terms, its structure is preserved. 
Finally, we asked whether going to subleading order in the $\beta/R$ expansion is possible within the membrane theory framework. We answered this question in the affirmative, and found that at subleading orders we have to include higher derivative corrections. 

We regard our work as an important demonstration of the versatility and robustness of the membrane theory. The rich applications of the theory include the demonstration of entropy inequalities that follow from it \cite{Bao:2018wwd}, its bit thread reformulation \cite{Agon:2019qgh}, application to R\'enyi entropies \cite{Zhou:2018myl} and logarithmic negativity \cite{Kudler-Flam:2019wtv}, and the exploration of shape dependence of entropy dynamics numerically \cite{toappear}. In the future, it would be interesting to generalize the holographic derivation of a membrane theory to other entanglement measures, such as R\'enyi entropies, negativity, and reflected entropy \cite{Dutta:2019gen}. Crucial stress tests of the theory would be to include bulk quantum   corrections in holography that are dual to $1/N$ corrections in the field theory and to test whether the membrane theory's predictions are correct for the entropy of a two interval subregion in chaotic spin chain numerics and arbitrary chaotic two-dimensional CFTs. If the membrane theory passes these feasible future tests, it would present an extremely strong case for the general applicability of the membrane theory for all chaotic systems in the hydrodynamic limit $R,T\gg t_\text{loc}$, where the local thermalization time $t_\text{loc}\sim\beta$ in strongly coupled theories, but could be significantly larger in weakly coupled (but chaotic) theories $t_\text{loc}\sim\beta/\lam$, where $\lam$ is some weak coupling constant. Instead of relying on hopefully representative examples to make the case for the membrane effective theory, it would be very desirable to present a derivation of it based on general principles that is applicable to all chaotic theories.

\section*{Acknowledgments} 
MM is supported by the Simons Center for Geometry and Physics. JV is supported by NSF award PHY-1620628.

\nocite{*}
\bibliographystyle{JHEP}
\bibliography{References}

\end{document}